\documentstyle[12pt,fps,amssymbo,cite,hlrzprep]{article}
   \newcommand{\nc}{\newcommand}
   \nc{\ds}{\displaystyle}
\makeatletter \if@double
   \nc{\be}{\begin{equation}\setlength{\linewidth}{159mm}}
   \nc{\bea}{\begin{eqnarray}\setlength{\linewidth}{159mm}}
\else
   \nc{\be}{\begin{equation}}
   \nc{\bea}{\begin{eqnarray}}
\fi\makeatother
   \nc{\bc}{\begin{center}}
   \nc{\eq}{\end{equation}}
   \nc{\ec}{\end{center}}
   \nc{\eqa}{\end{eqnarray}}
   \nc{\non}{\nonumber}
   \nc{\flift}{\rule[-0.5ex]{0ex}{2.5ex}}
   \nc{\tlift}{\rule[-1.0ex]{0ex}{3.0ex}}
   \nc{\slift}{\rule[-1.5ex]{0ex}{4.0ex}}
   \nc{\mlift}{\rule[-2.0ex]{0ex}{5.0ex}}
   \nc{\llift}{\rule[-2.5ex]{0ex}{6.0ex}}
   \nc{\ulift}{\rule[0ex]{0ex}{2.7ex}}
   \nc{\dlift}{\rule[-1.5ex]{0ex}{0ex}}
   \nc{\zr}{\phantom{0}}
   \nc{\phn}{\phantom}
   \nc{\al}{\alpha}
   \nc{\bt}{\beta}
   \nc{\gm}{\gamma}
   \nc{\dl}{\delta}
   \nc{\ep}{\epsilon}
   \nc{\vep}{\varepsilon}
   \nc{\zt}{\zeta}
   \nc{\th}{\theta}
   \nc{\kp}{\kappa}
   \nc{\lm}{\lambda}
   \nc{\sg}{\sigma}
   \nc{\vr}{\varphi}
   \nc{\om}{\omega}
   \nc{\Sg}{\Sigma}
   \nc{\Ps}{\Psi}
   \nc{\Ph}{\Phi}
   \nc{\Gm}{\Gamma}
   \nc{\cph}{\phi}                    
   \nc{\lph}{\Phi}                    
   \nc{\Mc}{{\cal M}}                 
   \nc{\Ml}{ M }                      
   \nc{\El}{ E }                      
   \nc{\mph}{\varphi}                 
   \def\itvec#1{\hspace*{0.15ex}
             \vec{{\hspace*{-0.08ex}#1\hspace*{0.08ex}}}
             \hspace*{0.15ex} }
   \def\emvec#1{\hspace*{0.15ex}
             \vec{{\hspace*{-0.25ex}#1\hspace*{0.25ex}}}
             \hspace*{0.15ex} }
   \def\rmvec#1{\hspace*{0.15ex}
             \vec{{\hspace*{0.10ex}#1\hspace*{-0.10ex}}}
             \hspace*{0.15ex} }
   \nc{\vp}{\itvec{p}}
   \nc{\vs}{\itvec{s}}
   \nc{\vcph}{\emvec{\cph}}
   \nc{\vlph}{\rmvec{\lph}}
   \nc{\vmph}{\itvec{\mph}}
   \nc{\vml}{\itvec{\Ml}}
   \nc{\vmc}{\emvec{\Mc}}
   \nc{\lra}{\longrightarrow}
   \nc{\lla}{\longleftarrow}
   \nc{\lag}{\left\langle\rule[-0.3ex]{0ex}{2.3ex}\hspace*{-0.1em}\right.}
   \nc{\rag}{\left.\rule[-0.3ex]{0ex}{2.3ex}\hspace*{-0.1em}\right\rangle}
   \nc{\tlag}{\left\langle\rule[-0.1ex]{0ex}{1.8ex}\right.\hspace*{-0.20em}}
   \nc{\trag}{\hspace*{-0.20em}\left.\rule[-0.1ex]{0ex}{1.8ex}\right\rangle}
   \nc{\tl}{\tilde}
   \nc{\p}{\partial}
   \def\smhat#1{\hspace*{0.15ex}
             \widehat{{#1}}
             \hspace*{0.15ex} }
   \nc{\phat}{\smhat{p}}
   \nc{\qhat}{\smhat{q}}
   \nc{\pqhat}{\smhat{(p-q)}}
   \nc{\phati}{\smhat{2p}}
   \nc{\pll}{\parallel}
   \nc{\Dsl}{D\!\!\!\!/}
   \nc{\dsl}{d\!\!\!\!/}
   \nc{\hmu}{\hat{\mu}}
   \nc{\Real}{\mbox{\rm Re }}
   \nc{\ZZ}{{\Bbb Z}}
   \nc{\RR}{{\Bbb R}}
   \nc{\NN}{{\Bbb N}}
   \nc{\lb}{\label}
   \def\arraystretch{1.0}
\begin{document}
\dateandnumber(June~1992)%
{HLRZ J\"ulich 92--35}%
{Aachen, PITHA 92--21}%
{\strut}%
\titleofpreprint%
{                   Scaling analysis                              }%
{       of the O(4)--symmetric $\Phi^4$--theory                   }%
{                   in the broken phase\footnotemark[1]           }%
{                                                                 }%
\listofauthors%
{             Meinulf G\"ockeler$^{1,2}$,
              Hans A.~Kastrup$^1$,
              Thomas Neuhaus$^3$,                                 }%
{             Frank Zimmermann$^{1,2}$                            }%
{                                                                 }%
\listofaddresses%
{\em      $^1$Institut f\"ur Theoretische Physik E, RWTH Aachen,
              W-5100 Aachen, Germany                              }%
{\em      $^2$HLRZ c/o KFA J\"ulich, P.O.Box 1913,
              W-5170 J\"ulich, Germany                            }%
{\em      $^3$Fakult\"at f\"ur Physik,
              Univ.~Bielefeld, W-4800 Bielefeld,
              Germany                                             }%
\abstractofpreprint{
  We study the $O(4)$--symmetric $ \Phi^4 $--theory in the scaling
region of the broken phase using the standard and a Symanzik improved
action with infinite bare self-coupling $\lm$.  A high precision Monte
Carlo simulation is performed by applying the reflection cluster
algorithm.  Employing the histogram method we analytically continue to a
sequence of values of the hopping parameter $\kp$ neighbouring the
actually simulated ones.  In the investigated vicinity of the critical
point $\kp_{c}$ finite volume effects affecting, e.g., the determination
of the field expectation value $\Sg$ and the mass $m_\sg$ of the
$\sg$--particle are very well described by $1$--loop renormalized
perturbation theory.  We carry out a  detailed scaling analysis
on a high level of precision.  Finally we discuss the upper bound on the
Higgs mass for both kinds of actions.
}
\footnoteoftitle{
\setcounter{footnote}{1}
\footnoteitem($\fnsymbol{footnote}$){%
Supported by the Deutsche Forschungsgemeinschaft}%
}
\clearpage
%
\section{Introduction}
  By now it seems to be well established (though not rigorously proven)
that the scalar sector of the standard model, the  $O(4)$--symmetric $
\Phi^4 $--theory, is trivial in the sense that all interactions vanish as
the cut--off is removed.  Therefore, an interacting theory can only be
obtained in the presence of a large but finite (momentum) cut--off, i.e.\
the model is to be regarded as an effective field theory.  Under the
assumption that the gauge interactions can be treated perturbatively,
this fact leads to an upper bound of the renormalized quartic coupling or
equivalently to an upper bound of the mass of the Higgs particle
\cite{daneu}.  This upper bound is of considerable interest, because it
indicates the energy scale where "physics beyond the standard model" has
to show up.  Its calculation, however, requires nonperturbative
techniques such as, e.g.\ , a Monte Carlo simulation.  Several analytical
and numerical studies lead to an upper bound on the Higgs mass
of about $650 \, GeV$ (see, e.g.,\ \cite{luwe,hmb,neub,pert}).

  Most of these investigations were performed on hypercubic lattices with
a nearest--neighbour action.  However, the value of the upper bound can
depend on the regularization used \cite{regu,f4}.  Therefore, a comparison
of different regularization schemes is of interest.

  In this paper, we study the  $O(4)$--symmetric $ \Phi^4 $--theory in
the broken phase (which is the relevant one for the standard model) by
means of Monte--Carlo simulations on hypercubic lattices.  We employ two
different discretizations of the continuum action:  the standard lattice
action with nearest--neighbour couplings only and an action which is
tree--level improved in the sense of Symanzik \cite{sym} and contains
couplings over a distance of two lattice spacings.

  We have computed the field expectation value $ \Sg $, the wave function
renormalization constant $ Z $ of the Goldstone bosons associated with
the spontaneous breakdown of the $O(4)$--symmetry and the mass $ m_\sg
$ of the Higgs particle, called $ \sg $--particle in the context of the
pure $ \Phi^4 $--model.  The renormalized field expectation value $ F $
(or the analogue of the pion decay constant) is then given by
 \be
                F = \frac{\ds \Sg}{\ds \sqrt{Z} } \; .
 \eq
The determination of these quantities will enable us to calculate the
upper bound for the Higgs mass and, in particular, to study the scaling
behaviour of the theory. Special attention is to be paid to the
logarithmic corrections of the mean field scaling laws.

  The numerical simulations can, of course, only be performed on finite
lattices.  Therefore, the results have to be extrapolated to infinite
volume.  This can be done reliably only on the basis of a thorough
understanding of the finite--size effects.
  In our case, the volume dependence comes from two sources:  First,
there are long--range correlations caused by the (almost) massless
Goldstone bosons.  Secondly, we have finite--size effects governed by the
$ \sg $--mass.  These will be sizeable unless the corresponding
correlation length is much smaller than the extent of the lattice.  For a
sufficiently large lattice the finite--size effects due to the Goldstone
bosons will dominate.
  One can then use chiral perturbation theory to extract infinite volume
results from the data obtained on finite lattices.  This method has been
applied very successfully in previous work \cite{chipt}.
  Here, however, we are particularly interested in the scaling behaviour
of the theory, so that we have to approach the critical surface as
closely as possible.
  This leads to $ \sg $--masses which are so small that on lattices of
the size we could handle the finite--size effects due to the $ \sg
$--particle cannot be neglected.  Hence chiral perturbation theory is no
longer appropriate to describe the volume dependence of our data and we
have to adopt another approach.

It is known that the $ \Phi^4 $--theory can be considered as weakly
interacting.
  Therefore, renormalized perturbation theory in the quartic coupling can
be applied inside the scaling region.
  Of course, this may be done in a finite as well as in an infinite
volume leading to a description of the finite--size effects within the
framework of standard perturbation theory \cite{pert}.
  Concerning these finite--size corrections, we shall see that the
effects due to the finite lattice spacing can be neglected in the region
of parameter values we have studied.

A further problem is created by the instability of the $ \sg $--particle:
It can decay into an even number of Goldstone bosons.
Although several methods have been developed \cite{resfin}
to deal with this situation
which do not rely on the particular model under consideration, a
direct determination of resonance parameters from numerical
data in four dimensions has not yet appeared in the literature.
In our model renormalized perturbation theory offers the
possibility to calculate the volume dependence of the relevant
quantities. In this way, an infinite--volume $ \sg $--mass will be obtained
corresponding to the real part of the pole of an appropriate propagator.
A more direct determination of the $ \sg $--mass, which does not take
recourse to perturbation theory, is however still highly desirable
and is currently under investigation.

  The plan of the paper is as follows:  Section 2 describes the model and
some tools applied in the subsequent analysis.  Section 3 deals with
perturbation theory in finite volume, which requires some care due to the
presence of the massless Goldstone bosons.  The formulas used for the
extraction of $ \Sg $, $ Z $, and $ m_\sg $ from our data are collected
in section 4. Details on the numerical simulations are given in section
5. In section 6 we exploit our results to analyze the scaling behaviour.
The upper bound to the Higgs mass is discussed in section 7. Section 8
contains our conclusions.  Finally, we have two appendices devoted to
some technical details.
   Preliminary accounts of our results have already been given in
ref.~\cite{paper}.
Note that the numbers presented there differ slightly from
those in the present paper due to an overlooked contribution
in one of the formulas used in the data analysis.
\section[The model on the lattice and its distribution functions]%
{The model on the lattice \\ and its distribution functions}
\subsection{The standard action and the Symanzik improved action}
We consider the $N$ component $O(N)$-invariant $ \Phi^4 $-theory on a
$d$-di\-men\-sio\-nal (hyper)cubic lattice of volume $ V \!=\! L^d $
($ L \in \NN $)
with periodic boundary conditions.
The numerical simulations will be done for $ N \!=\! 4 $, $ d \!=\! 4 $.
We choose a lattice regularized action of the form
\bea
\lb{eq:lact}
    S \{ \vlph ; J, \kp , \om \} & = & \!\!
         \sum_{ x\in \left( \ZZ_L \right)^d} \left\{ \mlift
         - \kp
         \sum_{ \mu = 1 }^{d} \left( \lph^\al (x) \lph^\al (x + \hmu) +
                                    \tlift
                                     \lph^\al (x) \lph^\al (x - \hmu)
                            \right) \right. \\
    \mbox{}                     &   & \left. \mlift
         \hphantom{\sum_{ x\in \left( \ZZ_L \right)^d} } \!\!
         - \kp \om
    \sum_{ \mu = 1 }^{d} \left( \lph^\al (x) \lph^\al (x + 2\, \hmu)  +
                                    \tlift
                                     \lph^\al (x) \lph^\al (x - 2 \, \hmu)
                            \right) \right. \non\\
    \mbox{}                     &   & \left. \mlift
         \hphantom{\sum_{ x\in \left( \ZZ_L \right)^d} } \!\!
     + \lm \left( \lph^\al (x) \lph^\al (x) - \tlift 1 \right)^{2}
     + \lph^\al (x) \lph^\al (x)
     - J \lph^N (x) \right\} \,, \non
\eqa
where $ \lph^\al (x) $, $ \al = 1, \ldots , N $, denotes the real
$O(N)$--vector field and $ \hmu $ is a unit vector pointing in the positive
$ \mu $--direction.
   Note that we have coupled the field to a symmetry breaking external
source $ J $ for later convenience.
Expectation values with respect to the (unnormalized) weight $ e^{-S} $ will
be denoted by $ \tlag \ldots \trag $. In the numerical simulations we have
put $\lm=\infty $ such that $ \vlph (x) $ is a vector of unit length.
For this value of $\lambda$ it is expected that the upper bound on the
Higgs mass takes its maximal value.

Introducing a dimensionful rescaled field $ \vcph   $ and a
corresponding source $ j $ according to
\bea
\lb{eq:phicont}
       \cph^\al (ax) &=& \sqrt{2 \kp} \sqrt{1 + 4 \, \om}  \; \lph^\al (x)
                        a^{1-d/2} \,, \\
\lb{eq:jcont}
       j            &=& \frac{1}{\sqrt{2 \kp}}
                        \frac{1}{\sqrt{1 + 4 \, \om}} \; J
                        a^{-d/2-1} \,,
\eqa
where $a$ is the lattice spacing, one obtains the action in continuum--like
parametrization (up to an irrelevant constant):
\bea
\lb{eq:actcont}
    S \{ \vcph ; j , \om \}  & \!=\! &
     a^d \!\! \sum_{ x\in \left( \ZZ_L \right)^d} \left\{ \mlift
         \frac{1}{1+4 \om} \frac{1}{2} \sum_{ \mu = 1 }^{d}
         \frac{\cph^\al (ax + a\hmu) - \cph^\al (ax)}{a} \;
         \frac{\cph^\al (ax + a\hmu) - \cph^\al (ax)}{a}
                                    \right. \non \\
    \mbox{}                     &   &
         \hphantom{a^d \!\! \sum_{ x\in \left( \ZZ_L \right)^d}}
         + \frac{4 \om}{1 + 4 \om} \, \frac{1}{2} \sum_{ \mu = 1 }^{d}
         \frac{\cph^\al (ax + 2a \hmu) - \cph^\al (ax)}{2a} \;
         \frac{\cph^\al (ax + 2a \hmu) - \cph^\al (ax)}{2a}
                                     \non \\
    \mbox{}                     &   &
         \hphantom{a^d \!\! \sum_{ x\in \left( \ZZ_L \right)^d}}
        - \frac{1}{2} m_0^2 \cph^\al (ax) \cph^\al (ax)
        + \frac{g_0}{4!} \left(\cph^\al (ax) \tlift\cph^\al (ax) \right)^2
        - j \cph^N (ax) \left. \mlift \right\}
\eqa
with
\bea
\lb{eq:masscont}
     a^2m_0^2 & = & \frac{1}{1+4 \om} \, \left[ \frac{2 \lm -1}{\kp}
                                        + 2 d \left( 1 + \om\right)
                                       \right] \,, \\
\lb{eq:coupcont}
   a^{4-d} g_0  & = & \frac{1}{(1+4 \om)^2}
   \, \frac{6 \lm}{\kp^2}
\,.
\eqa

We shall study the standard nearest--neighbour action, where the
coupling parameter $ \om $ is zero, and an action improved in the spirit
of Symanzik. In the latter case we choose $ \om $ such
that the ${\cal O}(a^2) $ corrections in the lattice dispersion
relation vanish.
Since the inverse free propagator in momentum space corresponding to the
kinetic term of the action (\ref{eq:lact}) is given by
\bea
       \lefteqn{  2 \kp \left\{
                        \left( 1 + 4 \om \right) \,
                         m^2 + \phat^2  +  \om \, \phati^2
                         \right\} } \\ \non \mbox{}
      & = &
            2 \kp \left( 1 + 4 \om \right) \,
                        \left\{ m^2 + \sum_{ \mu = 1 }^{d}
                           \left( p_\mu \right)^2 \,
                              \left[ 1 -
                                 \frac{\left( p_\mu a \right)^2}{12}
                                 \frac{\left( 1 + 16 \om \right)}{
                                 \left( 1 + 4 \om \right)}
                              \right]
                  \right\}
            + {\cal O} \left( a^4 \right),
\eqa
where
\be
            \phat^2  =  \frac{4}{a^2} \sum_{ \mu = 1 }^{d}
                           \sin^2 \left( \frac{p_\mu a}{2} \right) \,,
\eq
we have to put  $ \om = - \frac{1}{16} $.

  In the following we have $d=4$ and, for simplicity, we take the lattice
spacing $a$ equal to $1$, i.e.\ all quantities are expressed in lattice
units.

\subsection{Double distribution of the mean field and the energy}
\lb{sec:distri}
{}From a simulation of the model at certain coupling parameters $ \kp $ and
$J $ one can obtain results at neighbouring values of the parameters if
appropriate distributions have been measured \cite{ferswe}.
For the extrapolation in the external source $ J $ one needs the
distribution of the mean field
\be
       \frac{\ds 1}{\ds V} \sum_x  \vlph  ( x )
\eq
and for the extrapolation in $ \kp $ we have to measure the distribution
of the kinetic energy term
\bea
\lb{eq:o}
       \frac{\ds 1}{\ds V} \sum_x  {\cal O} ( x ) & = &
            \frac{\ds 1}{\ds V}  \sum_{ x,\mu } \left[ \left(
            \vlph  (x)  \cdot \vlph  (x + \hat{\mu}) +
            \vlph  (x)  \cdot \vlph  (x - \hat{\mu}) \right) \mlift
            \right. \\  \mbox{} &   &
            \hphantom{\frac{\ds 1}{\ds V}  \sum_{ x,\mu } }
            \left. \mlift + \om \left(
            \vlph  (x)  \cdot \vlph  (x + 2 \hat{\mu}) +
            \vlph  (x)  \cdot \vlph  (x - 2 \hat{\mu}) \right)
            \right] \,. \non
\eqa
Since we are mainly interested in small values of $ J $, we have simulated
the model at vanishing external source and a few values of $ \kp = \kp_0 $,
calculating the double distribution of the mean field and the energy
\bea
      \tl{Z}_{ \kp_0 } ( \Ml  , \El ) & = & \int\!
         [d\lph] \exp \left(-S \{  \vlph  ; J = 0, \kp_0, \om \}
                      \right) \, \\
         \mbox{  } & & \hphantom{ -S \{  \vlph  ; \kp_0, \om \}}
         \times \dl \! \left( \slift \right.
                             \El  - \frac{\ds 1}{\ds V } \sum_{x}
                        {\cal O} (x) \left. \slift
                        \right)
                \dl \! \left( \slift \right.
                             \vml - \frac{\ds 1}{\ds V} \sum_x \vlph ( x )
                        \left. \slift \right) \, . \non
\eqa
 Due to the $O(N)$--symmetry at $ J \!=\! 0 $ it  depends only on the
absolute value $ \Ml \!=\! |\vml| $ of the mean field and on the energy
$\El$.

For the distribution of the mean field at a neighbouring $ \kp \neq
\kp_0 $ we have
\be
      \bar{Z}_{\kp} (\Ml) =
         \int\limits_{0}^{\infty} \!\! d \El
         \exp \left( V (\kp - \tlift \kp_0) \El \right)
         \tl{Z}_{ \kp_0 } ( \Ml  , \El ) \,.
\eq
 If we knew the double distribution $ \tl{Z}_{ \kp_0 } ( \Ml  , \El ) $
 of $ \El $ and $ \Ml $ at a simulation point  $ J \!=\! 0 $,
 $\kp\!=\!\kp_0$
with infinite accuracy, we could compute  the mean field distribution
$\bar{Z}_{\kp} (\Ml)$  at any value of $ \kp$.
Since in our numerical  simulations  statistics is, of course, limited,
we are using a discrete  form of the double distribution with an array
of $ 400^2 $ bins, and it is only possible
to extrapolate (or interpolate) to neighbouring values of $ \kp $ where
one has enough overlap between the functions
$ \exp \left( V (\kp - \kp_0) \El \right)$ and
$ \tl{Z}_{ \kp_0 } ( \Ml  , \El )  $.

In the presence of a symmetry breaking external source $ J $ the
partition function $ Z(J ,\kp) $ can be written as
an ordinary integral over the mean field $ \vml $ and the energy $ \El $
\bea
     Z(J, \kp) & = &  \int   [d \lph]
                   \exp \left(- S\{ \vlph ; J , \kp , \om \}
                        \right) \\
       \mbox{} & = &  \int \! d^N \!\! \Ml \int\limits_{0}^{\infty}\!\! d\El
                   \exp \left( \tlift V \, J \, \Ml_N \right) \,
                  \exp \left( \tlift V \, (\kp - \kp_0) \El
                  \hphantom{\rule[0ex]{0ex}{2.5ex}} \right) \,
                  \tl{Z}_{ \kp_0 } ( \Ml  , \El ) \non \\
       \mbox{} & = & \int \! d^N \!\! \Ml
                  \exp \left( \tlift V \, J \, \Ml_N  \right) \,
                  \bar{Z}_{ \kp } ( \Ml ) \,. \non
\eqa
We could now carry out the angular integration. However, in order to monitor
the distribution of the component $ \Ml_N$, which is coupled to the source
$ J $, we prefer to do only $ N-2 $ of these integrations.
 With
\be
      g(\Ml,\Ml_N) = \Theta(\Ml- | \Ml_N | ) \,  \Ml^{ N-2 }
                 \left( 1 - \left( \frac{\Ml_N}{\Ml} \right)^2 \right)
                                   ^{ {(N-3)} / {2} }
             \frac{2 \pi^\frac{N-1}{2}}{\Gm\left(\frac{N-1}{2}\right)}
\eq
we get for
the partition function
\be
      Z(J,\kp) = \! \int\limits_{0}^{\infty}  \!\! d\Ml \!
                  \int\limits_{-\infty}^{\infty} \!\! d\Ml_N \!
                  \int\limits_{0}^{\infty} \! d\El \, g(\Ml,\Ml_N)
                  \exp \left(\tlift V J \Ml_N \right)
                  \exp \left(\tlift V  (\kp - \kp_0) \El \right)
                  \! \tl{Z}_{ \kp_0 } ( \Ml  , \El ) .
\eq
This is useful for defining the probabilities
for the absolute value of the mean field $ \Ml $,
for the component $ \Ml_N $ of the mean field parallel to the external
source,
and for the energy $ \El $
to be contained in $ d\Ml $, $ d\Ml_N $ and $ d\El $, respectively:
\bea
      d\bar{P}_{ J ,\kp } (\Ml) & = &
           \frac{1}{\ds Z(J,\kp) }
           \left( \; \int\limits_{-\infty}^{\infty} \!
           d\Ml_N  g(\Ml,\Ml_N) \exp \left(V J \tlift \Ml_N \right)
           \right)
           \bar{Z}_{ \kp } ( \Ml ) d\Ml \,, \non \\
      dP_{ J ,\kp } (\Ml_N)     & = &
           \frac{1}{\ds Z(J,\kp) }
           \left( \int\limits_{0}^{\infty} \!
           d\Ml \bar{Z}_{\kp}(\Ml) g(\Ml,\Ml_N) \right)
           \exp \left( \tlift V \, J \, \Ml_N  \right)
           d\Ml_N \,, \\
      d\tl{P}_{ J ,\kp } (\El) & = &
           \frac{1}{\ds Z( J,\kp) }
           \left(\int\limits_{0}^{\infty} \!
           d\Ml \int\limits_{-\infty}^{\infty} \! d\Ml_N
           \exp \left(\tlift V \, J \, \Ml_N \right)
           \tl{Z}_{ \kp_0 } ( \Ml  , \El ) \, g(\Ml,\Ml_N)  \right)
           \non \\
      \mbox{}                &   & \times
            \exp \left(\tlift  V \, (\kp - \kp_0) \, \El \right)  d\El
            \,. \non
\eqa

The meaning of $ g(\Ml,\Ml_N)$ becomes clear if we suppose a delta
distribution $ \bar{Z} \sim \dl \left(\bar{\Ml} -\Ml\right) $ for the
absolute value  $ \Ml $ of the mean field.
 In this case the  distribution of one component $ \Ml_\al $ is proportional
to $ g(\bar{\Ml},\Ml_\al) $ if $ J \!=\! 0$.

In this way we get a relatively smooth distribution of the component
$ \Ml_N $ living in the interval $ \left[ -\Ml,+\Ml \right] $ from the
peaked distribution of the absolute value $ \Ml $ of the mean field.
  A direct measurement of the distribution of $ \Ml_N $ would give a less
smooth result, because it is much more efficient to divide the smaller
interval over which $\Ml$ varies into bins than the larger one, which is
covered by $\Ml_N$.

  We are now able to describe the interesting expectation values in a
very compact way, for example the expectation value of the component of
the field pointing in the direction of the external source
\be
      \lag \frac{1}{V} \sum_{x} \lph^{N} (x) \rag_{ J,V,\kp}
            = \tlag \Ml_N \trag_{ J,V,\kp}
            = \int\limits_{-\infty}^{\infty} \! dP_{ J,\kp } (\Ml_N) \Ml_N
   \lb{eq:distproj}
\eq
and the expectation value of the absolute value of
the mean field
\be
      \lag \left| \slift \right.
         \frac{1}{V} \sum_{x} \vlph (x) \left.
         \slift \right| \rag_{ J,V,\kp}
            = \tlag \Ml \trag_{ J,V,\kp}
            = \int\limits_{0}^{\infty}\! d\bar{P}_{ J ,\kp } (\Ml) \Ml \,,
   \lb{eq:distabs}
\eq
which were numerically calculated for $ \kp \!=\! \kp_0 $
and small $ J $ as well as for $ J \!=\! 0 $ and $ \kp $ values in the
neighbourhood of $ \kp_0 $.

\section{Perturbation theory and finite--size effects}
\lb{sec:finsiz}
In this section we derive some formulas which will be used later to
extract the  vacuum expectation value $\Sg$ of the unrenormalized field,
 the Goldstone boson wave  function renormalization constant $ Z $, and
the mass $m_\sg$  of the $\sg$--particle from the numerical data.
In contrast to previous experience \cite{chipt},
the application of chiral perturbation
theory turned out to be unsuccessful  in the range of
$ \kp $-values analysed here, which are very close to the phase transition.
On lattices of the size we could handle, the product $ m_\sg L $
is too small to suppress the finite--size effects due to the $ \sg $--mass,
so that one main condition for the application of chiral perturbation theory
is not satisfied.
Hence we used formulas derived from $1$--loop perturbation theory in the
renormalized coupling $ g_R $.
To account for finite--size effects we did the perturbation theory in
a finite volume combined with a saddle point approximation.
We perform our perturbative calculations on the lattice.
 The renormalization constants will be calculated in the infinite volume
 according to the renormalization conditions of L\"uscher and Weisz.
 Hence our renormalized parameters refer to the infinite volume model.
 In the limit $ L \!\to\! \infty $ the perturbative results take the form
predicted  by chiral perturbation theory with the coupling constants of
the chiral  effective Lagrangean expressed in terms of the
renormalized parameters.
Note that we shall employ the continuum normalization of fields
(compare eqs.\ (\ref{eq:phicont}) -- (\ref{eq:coupcont})) throughout this
section.

In order to remove the infrared divergences of the Goldstone propagator
one can introduce in a finite volume an external source term
    $ j \sum\limits_{x} \cph^N (x)$,
which equips the Goldstone bosons with a mass
    $ m_\pi \sim \sqrt{j} $.
The source $j$ can be sent to zero only after the infinite--volume limit is
taken. In a second method, which we choose here,
    the mean field
    $ \frac{1}{V} \sum\limits_{x} \vcph  (x) $
    (the momentum zero mode)
    is treated separately as a collective variable,
so that the limit $j\!\to\! 0$ does not cause problems even in the finite
volume.

\subsection{Perturbation theory at finite volume}
\subsubsection{Separation of the momentum zero modes }
We split off the zero momentum part of the functional
integral for the partition function
\bea
         Z(j) & = & \int\! \left[ d\cph  \right]
              \exp \left(- S \{
                    \vcph ; j , \om \} \right)
                 =   \int \! d^N \!\! \Mc \int \!
                     \left[ d \mph  \right]
              \exp \left(- S
              \{ \vmph  + \vmc; j  , \om \} \right)  \non \\
      \mbox{} & = & \int \! d^N \!\!  \Mc \; \tl{\cal Z} (j,\vmc)
\lb{eq:parfct}
\eqa
by decomposing the field into its components with zero and nonzero momentum,
respectively:
\be
      \vcph  (x) = \vmph  (x) + \vmc  \,, \quad \quad
             \vmc = \frac{1}{V} \sum_{x} \vcph  (x)
                            \,, \quad \quad
                       \sum_{x} \vmph (x) = 0 \,.
\eq
The length of the mean field $ \vmc $ is denoted by $ \Mc $.

In terms of the constraint functional integral
 \be
      \tl{\cal Z} (j,\vmc) = \int\!
               \left[ d \cph \right] \,
               \dl \! \left( \slift \right.
                        \vmc - \frac{\ds 1}{\ds V} \sum_x \vcph  ( x )
                       \left. \slift \right) \,
                       \exp \left(- S \{ \vcph ; j , \om \} \right)
\eq
the constraint effective potential $ {\cal U } (\Mc) $
is given by
\be
      \tl{\cal Z} (j,\vmc) =
               \exp \left\{ \slift -V {\cal U} (\Mc) +V j \Mc_N \right\}
\lb{eq:zpot}
\eq
up to a normalization constant.
Note that the volume dependence of $ {\cal U} (\Mc) $ is not displayed
explicitly.

In eq.~(\ref{eq:parfct}) the angular integral over the direction  of
$ \vmc $ can  be carried out
since it only enters through the source term $ \exp(V j \Mc_N) $:
\bea
      Z(j)   & = &  2 \pi^{\frac{N}{2}}
                     \int\limits_{0}^{\infty}\!
                     d\Mc \, \Mc^{N-1} Y_N (V j \Mc)
                     \exp \left( \tlift  - V {\cal U}(\Mc)  \right) \,, \\
      Y_N(x) & = &  \left(\frac{x}{2} \right)^{-\nu} \, I_\nu (x) \,,
                    \quad \nu = \frac{N}{2} - 1 \,.
\eqa
 $ I_{\nu}(x) $ are the well--known modified Bessel functions.
In our later calculations  we sometimes use the relation
\be
         x Y_N''(x) = - (N-1) Y_N'(x) + x Y_N(x) \,,
\eq
without further reference.

In addition to the mean field $ \vmc $, we investigate the two--point
functions of the projected field
\be
       \cph_{\pll}  (x) =
                          \frac{\sum_{\al} \cph^{\al} (x) \Mc_{\al}}{\Mc} \,,
\eq
 and of the transverse field
\be
       \cph^{\al}_{\perp} (x) =
                          \cph^{\al} (x) - \cph_{\pll} (x) \,
                          \frac{\ds \Mc_{\al}}{\Mc}  \,.
\eq
In an approximate sense we may think of $ \cph_{\pll}(x) $ as the
$ \sg $--field and of  $ \cph_{\perp}^{\al}(x) $ as the field of the
Goldstone bosons, although the correlation functions of $ \cph_{\pll} $
receive (nonleading) contributions also from the Goldstone bosons,
as can be seen e.g.\ in perturbation theory.
A similar caveat applies to the interpretation of $ \cph_{\perp} ^{\al}$.
Furthermore one should note that $ \cph_{\pll}(x) $ and
$ \cph_{\perp}^{\al}(x) $ are nonlocal: Through the mean field $ \vmc $
they depend on the whole field configuration.

For later purposes it is convenient to introduce the so--called constraint
correlation functions $ \tlag \cph^{\al} (x)  \cph^{\bt} (y) \trag_{\vmc}  $
defined by the following functional integral at $ j=0 $ \cite{goeleu}:
\bea
      \lag \cph^{\al} (x)  \cph^{\bt} (y) \rag_{\vmc} & = &
       \frac{\ds 1}{\ds \tl{\cal Z} (0,\vmc ) }
               \int\! \left[ d \cph \right]
               \dl \! \left( \slift \right.
                        \vmc - \frac{\ds 1}{\ds V} \sum_x \vcph ( x )
                      \left. \slift \right)  \\
       \mbox{} & & \hphantom{ \frac{\ds 1}{\ds \tl{\cal Z} (0,\vmc ) }
                             \int\! } \times
                       \exp \left(- S \{ \vcph ; j = 0, \om
                          \}\right)
                       \cph^{\al} (x)  \cph^{\bt} (y) \,. \non
\eqa
The two--point function of the field $ \vcph  (x) $ in the presence of
an external source $ j $ can then be written as
\be
      \lag \cph^{\al} (x)  \cph^{\bt} (y) \rag_{j} =
                \frac{\ds 1}{\ds Z(j)} \int \! d^N \!\! \Mc
                \, \tl{\cal Z} (j,\vmc) \,
                \lag \cph^{\al} (x)  \cph^{\bt} (y) \rag_{\vmc}\,.
\eq
Due to the $O(N)$--symmetry of the action with $ j=0 $ we have
the decomposition
\be
      \lag \cph^{\al} (x)  \cph^{\bt} (y) \rag_{\vmc} =
                         \frac{\ds \Mc_{\al} \Mc_{\bt} }{\ds \Mc^2}
                         \tl{G}_{\pll} ( x-y, \Mc ) +
                         \left( \dl_{\al \bt} -
                         \frac{\ds \Mc_{\al} \Mc_{\bt} }{\ds \Mc^2} \right)
                         \tl{G}_{\perp} ( x-y, \Mc ) \,.
\eq
The correlation functions of $ \cph_{\pll} $ and $ \vcph _{\perp} $
can now be expressed in terms of $ \tl{G}_{\pll} $ and $ \tl{G}_{\perp} $:
\bea
      \lag \cph_{\pll} (x)  \cph_{\pll} (y) \rag_{j} & = &
                \frac{\ds 1}{\ds Z(j)} \int \! d^N \!\! \Mc
                \, \tl{\cal Z} (j,\vmc) \, \tl{G}_{\pll} ( x-y, \Mc) \,, \\
      \lag \cph^{\al}_{\perp} (x)  \cph^{\bt}_{\perp} (y) \rag_{j} & = &
                \frac{\ds 1}{\ds Z(j)} \int \! d^N \!\! \Mc
                \, \tl{\cal Z} (j,\vmc) \,
                \left( \dl_{\al \bt}
                     - \frac{\ds \Mc_{\al} \Mc_{\bt} }{\ds \Mc^2}
                \right)
                \tl{G}_{\perp} ( x-y, \Mc) \,. \non
\eqa
Carrying out the angular integrations we obtain
\bea
      \lag \cph_{\pll} (x)  \cph_{\pll} (y) \rag_{j} & = &
                \frac{\ds 2\pi^{N/2}  }{\ds Z(j)}
                \int\limits_{0}^{\infty} \!
                d \Mc \, \Mc^{N-1} Y_N ( Vj\Mc ) \\
             \mbox{} & & \hphantom{ \frac{\ds 1}{\ds Z(J)} 2\pi^{N/2}
                                   \int\limits_{0}^{\infty} \!}
                \times \exp \left( - \tlift V {\cal U} (\Mc) \right)
                \tl{G}_{\pll} ( x-y, \Mc) \,, \non \\
      \lag \cph^{\al}_{\perp} (x)  \cph^{\bt}_{\perp} (y) \rag_{j} & = & 0
                \hspace*{63mm}
                \mbox{for} \quad \al \neq \bt \,,  \\
      \lag \cph^{\al}_{\perp} (x)  \cph^{\al}_{\perp} (y) \rag_{j} & = &
                \frac{\ds 2\pi^{N/2}  }{\ds Z(j)}
                \int\limits_{0}^{\infty} \! d \Mc \, \Mc^{N-1}
                \frac{\ds (N-2) Y_N ( Vj\Mc ) + Y_N''( Vj\Mc )}{\ds N-1 } \\
                \mbox{} & &
                \times \exp \left( - \tlift V {\cal U} (\Mc) \right)
                \tl{G}_{\perp} ( x-y, \Mc) ,
                \quad \mbox{for} \quad \al = 1, \ldots, N-1, \non \\
      \lag \cph^{N}_{\perp} (x)  \cph^{N}_{\perp} (y) \rag_{j} & = &
                \frac{\ds 2\pi^{N/2}  }{\ds Z(j)}
                \int\limits_{0}^{\infty} \! d \Mc \, \Mc^{N-1}
                \left[ Y_N ( Vj\Mc ) \tlift - Y_N'' ( Vj\Mc ) \right] \\
             \mbox{} & & \hphantom{ \frac{\ds 1}{\ds Z(j)} 2\pi^{N/2}
                                   \int\limits_{0}^{\infty} \!}
                \times \exp \left( \tlift - V {\cal U} (\Mc) \right)
                \tl{G}_{\perp} ( x-y, \Mc) \,. \non
\eqa

\subsubsection{Saddle point expansion}
If we consider  the constraint effective potential $ {\cal U} (\Mc) $
 to be known  (e.g.\ in perturbation theory)
 $ Z(j) $ can be approximated  by a saddle point expansion
 around the minimum $ \Mc_0 $ of $ {\cal U} (\Mc) $ as $ V \!\to\! \infty $.
 Introducing the notation
\be
      A_q = 2 \pi^{\frac{N}{2}}
               \int\limits_{0}^{\infty} \! d\Mc \; \Mc^{q} \; Y_N(V j \Mc)
               \exp \left(- \tlift  V {\cal U}(\Mc) \right)
\eq
 we can write $ Z(j) = A_{N-1} $, and the saddle point approximation
 yields (see appendix \ref{app:saddle} for a review of this method)
 for $ V \!\to\! \infty $ with $ V j $ fixed
\bea
      A_q       & = &  2 \pi^{\frac{N}{2}}
                       \exp \left( \tlift - V {\cal U} (\Mc_0 ) \right)\;\;
                          \left( b_0 (x,q) \, \mlift \right.
                                              \sqrt{\frac{2\pi}{V}}
                               + b_2 (x,q) \, \sqrt{\frac{2\pi}{V^3}}
                               + \left. \mlift \cdots
                          \right)_{x=V j \Mc_0} \,, \non \\
      b_0 (x,q) & = &  \Mc_0^{q}  \, Y_N(x) \, a_1 \slift \,,\\
      b_2 (x,q) & = &  \Mc_0^{q-2}
                       \left\{
                          \left[ \slift \right. q(q-1) \frac{a_1^3}{2}
                                 + 3 q a_1 a_2 \Mc_0 + 3 a_3 \Mc_0^2
                                 \left. \slift
                          \right] Y_N(x) \mlift  \right.  \non \\
      \mbox{}   &   &  \hphantom{ \Mc_0^{q-2} \left\{ \mlift \right.}
                        \quad + \left. \mlift
                          \left[ \slift 3 a_1 a_2 \Mc_0
                                 + q a_1^3
                          \right] x Y'_N(x)
                          + \frac{a_1^3}{2} \, x^2 \, Y''_N(x)
                       \right\} \non \,.
\eqa
The coefficients $ a_i $ only depend on the effective potential
$ {\cal U}(\Mc) $ and its derivatives at the minimum $ \Mc_0 $, and
are determined  in appendix \ref{app:saddle} for $ i=1,2,3 $.
Here we need $ a_1 $ and $ a_2 $ only:
\be
    a_1 = \left( \tlift {\cal U}''(\Mc_0) \right)^{-{1} / {2}}
          \quad, \qquad
    a_2 = - \frac{1}{6} \frac{{\cal U}^{(3)}(\Mc_0)}{
                        \left( \tlift {\cal U}''(\Mc_0) \right)^{2}} \,.
\makeatletter
\if@double\else
\pagebreak[4]
\fi\makeatother
\eq
In particular, we are interested in  the expectation value of the field
 $ \cph^N $
\bea
\lb{eq:mnj}
      \tlag\Mc_N\trag_{j ,V} \! & = & \! \lag \frac{1}{V} \sum_{x}^{}
                                    \cph^N \rag_{j , V}
                        =        \frac{ 1}{ V} \frac{\ds \p}{\ds \p j}
                                 \ln Z(j) \\
                      & = & \!   \frac{x}{j V}
                            \left\{\frac{1}{b_0(x,N-1)} \frac{\p}{\p x}
                                     b_0 (x,N-1) \llift
                                 + \frac{1}{V} \frac{\p}{\p x}
                                 \left(\frac{b_2(x,N-1)}{b_0(x,N-1)}\right)
                            \right\}_{x=V j \Mc_0} \non   \\
                      & = &      \frac{1}{j V}
                            \left( \frac{u Y_N'(u)}{Y_N(u))} \right)
                            + j a_1^2  \,, \non \\
              \mbox{} &   &
                              u = Vj\Mc_0
                            \left[ 1 + \frac{1}{V \Mc_0^2}
                              \left( 3 a_2 \Mc_0 + \frac{N-1}{2} a_1^2 \right)
                            \right]   \,, \non
\eqa
and in the expectation value of the  $ k $th power of $ \Mc $
\bea
\lb{eq:mnk}
   \tlag\Mc^k\trag_{j , V} \! & = & \! \! \lag
                           \left| \slift \right.
                           \frac{1}{V} \sum_{x}^{} \vcph (x)
                           \left. \slift \right|^k \rag_{ j,V }
                      =   \frac{\ds A_{ N+k-1 }}{\ds A_{N-1}} \\
                    & = & \! \Mc_0^k \,
                          \left\{ 1 + \llift \frac{1}{V}
                             \left( \frac{b_2(x,N+k-1)}{b_0(x,N+k-1)}
                                  - \frac{b_2(x,N-1)}{b_0(x,N-1)}
                             \right)
                          \right\}_{x=Vj\Mc_0} \non \\
                    & = & \! \Mc_0^k \,
                          \left\{ 1 + \frac{k}{V \Mc_0^2}
                             \left[ \slift \right.
                                 3 a_2 \Mc_0 + \frac{(2N+k-3)}{2} a_1^2
                                 +  \left. \frac{xY_N'(x)}{Y_N(x)}
                                    \right|_{x=Vj\Mc_0} \!\! a_1^2
                             \left. \slift \right]
                          \right\} \,. \non
\eqa
 Note that  for vanishing external source the $ j $--dependent
term in the last equation vanishes.

We should remark that the width of the mean field distribution
is simply given by the square root of $ \chi / V $, where the
susceptibility $ \chi $ is defined by
\be
      \chi = V \left[ \tlag\Mc^2\trag_{j , V}
              - \left(\tlag\Mc\trag_{j , V} \right)^2 \, \right] =
               a_1^2   =
               \left(\slift  {\cal U}'' (\Mc_0) \right)^{-1} \,.
\lb{eq:chidef}
\eq
In the order considered here
it does not depend on the size of the external source $ j $.

For the correlation functions of the projected field
 $  \cph_{\pll}  (x) $ and of the transverse field
 $\cph^{\al}_{\perp} (x) $ at $ j=0 $
the saddle point expansion yields to lowest order
\bea
      \lag \cph_{\pll} (x)  \cph_{\pll} (y) \rag_{j=0} & = &
                 \tl{G}_{\pll} ( x-y, \Mc_0) \,, \\
      \lag \cph^{\al}_{\perp} (x)  \cph^{\bt}_{\perp} (y) \rag_{j=0} & = &
                \dl_{\al \bt} \frac{\ds N-1}{\ds N}
                \tl{G}_{\perp} ( x-y, \Mc_0) \,. \non
\eqa
So we are left with calculating perturbatively the constraint effective
potential
$ {\cal U} (\Mc) $, its minimum $ \Mc_0 $, and its derivatives
$ {\cal U}''(\Mc_0)$, $ {\cal U}^{(3)} (\Mc_0) $ at the minimum,
as well as the constraint correlation functions.
 We shall restrict ourselves to $1$-loop order.

\subsubsection{1--loop perturbation theory}
At $1$-loop order we find for the standard action ($ \om \!=\! 0 $)
\bea
   \ds {\cal U} (\Mc) & = & - \frac{\ds m_0^2}{\ds 2} \Mc^2
                            + \frac{\ds g_0}{\ds 4!} \Mc^4
                           + \frac{1}{2V} \!
                        \sum_{ p\in \left( \frac{2\pi}{L} \ZZ_L \right)^4
                              }^{\hfill\prime\quad}
                        \ln \left(\phat^2 - m_0^2 + \frac{g_0}{2} \Mc^2
                            \right)
        \\ \mbox{ } &  &
                  \hphantom{- \frac{\ds m_0^2}{\ds 2} \Mc^2
                            + \frac{\ds g_0}{\ds 4!} \Mc^4 }
                      + \frac{N-1}{2V} \!
                        \sum_{ p\in \left( \frac{2\pi}{L} \ZZ_L \right)^4
                              }^{\hfill\prime\quad}
                        \ln \left( \phat^2 - m_0^2 + \frac{g_0}{6} \Mc^2
                            \right)
        \non
\eqa
The prime at the sum indicates that the momentum $p=0$ is left out.
For the minimum $ \Mc_0 $ we get
\be
      \Mc_0  =  \sqrt{ \frac{\ds 6 m_0^2}{\ds g_0} }
               \left( 1 - \frac{3}{2}\, \frac{g_0}{6 m_0^2} \,
                                        \bar{G}_1 (2m_0^2)
                        - \frac{N-1}{2} \,\frac{g_0}{6 m_0^2}\,
                                        \bar{G}_1 (0)
                        + {\cal O}(g_0^2)
               \right) \non
\eq
where we have introduced
\be
      \bar{G}_r (m^2) = \frac{1}{L^4}
                        \sum_{ p\in \left( \frac{2\pi}{L} \ZZ_L \right)^4
                              }^{\hfill\prime\quad}
                        \left( \phat^2 + m^2 \right)^{-r} \,.
\eq

After calculating the second and third derivative of the effective potential
we can evaluate the quantities
\bea
\lb{eq:a1}
        \chi =
        a_1^2 \! & = \! & \left( \tlift {\cal U}'' (\Mc_0) \right)^{-1}
      \\ \mbox{} & = \! & \frac{3}{g_0 \, \Mc_0^2}
                     \left\{ \slift \right.
                           1 + \frac{g_0}{6}
                           \left( 9 \, \bar{G}_2 (2m_0^2)
                            + (N-1) \, \bar{G}_2 (0) \right)
                            + {\cal O}(g_0^2)
                     \left. \slift \right\} \,, \non \\
\lb{eq:a2}
    3 a_2 \Mc_0 \! & = \! & - \frac{1}{2} \,
         \frac{ \Mc_0\; {\cal U}^{(3)} (\Mc_0) }
         {\left( \tlift {\cal U}'' (\Mc_0)\right)^2} \\
           \mbox{} & = \! & - \frac{3}{2} a_1^2 - \frac{g_0}{6}
                        \left( 27 \bar{G}_3(2m_0^2)
                            + (N-1) \bar{G}_3(0) \right)
                            + {\cal O}(g_0^2) \,, \non
\eqa
 which enter in eqs. (\ref{eq:mnj}) and (\ref{eq:mnk}).

   For the correlation function of the projected field we obtain
-- in combination with the saddle point expansion to lowest order --
\bea
     \lag \cph_{\pll} (x)  \cph_{\pll} (0) \rag_{j=0}
             & = & \Mc_0^2 + G ( 2 m_0^2, x ) \\
     \mbox{} &   & \mbox{} - \frac{g_0}{3}
                               \left(
                                     3 G ( 2 m_0^2, 0 )
                                      + (N-1) G ( 0 , 0 )
                               \right) \frac{\p}{\p m^2}
                               \left. G ( m^2 , x )
                               \right|_{m^2 = 2 m_0^2} \non \\
    \mbox{} &   & + m_0^2 \frac{g_0}{3}
                          \sum_{x', \, y'}
                          G ( 2 m_0^2 , x - x' ) \;
                          G ( 2 m_0^2 , y' )  \non \\
    \mbox{} &   & \hphantom{ m_0^2 \frac{g_0}{3}
                          } \times
                          \left(   9 \,G ( 2 m_0^2, x' - y' )^2
                                 + (N-1) \,G ( 0 , x' - y' )^2
                          \right)
                        + {\cal O}(g_0^2) \,, \non
\eqa
where
\be
      G (m^2, x) = \frac{1}{L^4}
                        \sum_{ p\in \left( \frac{2\pi}{L} \ZZ_L \right)^4
                              }^{\hfill\prime\quad}
                        \frac{\ds e^{ipx}}{\ds  \phat^2 + m^2 } \,.
\eq
%
\subsubsection{Renormalization}
\lb{sec:renorm}
All these perturbative relations require of course renormalization.
We calculate the renormalization constants for vanishing external
source on an infinite lattice for the standard nearest--neighbour action
applying the same renormalization conditions as L\"uscher and Weisz
\cite{luwe}.
For the Goldstone boson wave function renormalization constant $ Z $
we get
\be
         Z = 1 + \frac{2}{3} g_0 m_0^2
               \int\limits_{-\pi}^{\pi} \!
               \frac{d^4 p}{\left(2 \pi\right)^4} \,
               \frac{1}{\phat^2} \left[
                   - \frac{\cos  p_1 }{
                   \left( \phat^2 + 2 m_0^2 \right)^2 }
               + 4 \frac{\sin^2  p_1 }{
                   \left( \phat^2 + 2 m_0^2 \right)^3 }
                   \right]
               + {\cal O} \left( g_0^2 \right) \,.
\eq
For the renormalized $ \sg $--mass we obtain
\bea
      m_R & = & \mbox{\rm arcosh } \left( 1 + m_0^2 \right)
               - \frac{g_0}{12 m_0 \sqrt{2 + m_0^2} }
                 \; \Real \!\! \left[
                     6 \hat{I}_0 \left( 2 m_0^2 \right)
                  +  2 (N-1) \hat{I}_0 \left( 0 \right)
                     \mlift   \right. \non \\
      \mbox{} & & \left. \mlift \hspace*{15mm}
                  +  2 m_0^2 \left(
                              9 \hat{I}_1 \left( 2 m_0^2, p \right)
                            + (N-1) \hat{I}_1 \left( 0, p \right)
                             \right)
                    \right]_{p=\bar{p} } \!
                  + {\cal O} \left( g_0^2 \right) \,, \\
        \bar{p} & = & \left( 0,0,0,i \,
        \mbox{\rm arcosh} ( 1 + m_0^2 ) - \ep  \right)
\eqa
where
\bea
        \hat{I}_0 \left( m^2 \right) & = &
               \int\limits_{-\pi}^{\pi} \!
               \frac{d^4 p}{\left(2 \pi\right)^4}
               \frac{1}{\phat^2 + m^2}  \\
        \hat{I}_1 \left( m^2, p \right) & = &
               \int\limits_{-\pi}^{\pi} \!
               \frac{d^4 q}{\left(2 \pi\right)^4}
               \frac{1}{\left( \flift \right.
                        \qhat^2 + m^2
                        \left. \flift \right)
                        \left( \flift \right.
                        \pqhat^2 + m^2
                        \left. \flift \right)}
\eqa
and $ \ep > 0 $ is infinitesimal.
The renormalized $\sg$--mass $m_R$ as defined by L\"uscher and Weisz
\cite{luwe} is equal to the true particle mass
$m_\sg$ corresponding to the real part of the pole of the propagator
up to higher order corrections.

Note that even at tree level the renormalized and bare mass
agree only up to scaling violation terms.
An analogous statement holds for
the renormalized coupling constant defined as
\be
         g_R = 3 Z \frac{m_R^2}{\Sg^2}
             = 3   \frac{m_R^2}{F^2} \, ,
\eq
where $ \Sg \!=\! \lim\limits_{j \to 0} \tlag\cph^N\trag_{j}$
is the field expectation value for
vanishing source.

In the continuum limit we find
\bea
\lb{eq:mineff}
         \Mc_0 & = & \Sg
                   \left( 1 + B_1 ( m_{R} L) \frac{Z}{\Sg^2 L^2}
                            + {\cal O}( g_R^2 )
                   \right) \,, \\
\lb{eq:chipert}
                \chi
              & = & \frac{1}{m_{R}^2}
                    \frac{\Sg^2}{\Mc_0^2}
                          \left( 1 + \frac{m_R^2 Z}{2 \Sg^2} B_2 ( m_{R} L)
                          + {\cal O}( g_R^2 )
                          \right) \,, \\
\lb{eq:a2m}
    3 a_2 \Mc_0 \! & = \! &
                     - \frac{3}{2} \chi  - \frac{m_R^2 Z}{2\Sg^2} L^2
                                          B_3(m_R L)
                            + {\cal O}(g_R^2) \,,
\eqa
where
\bea
       \lb{eq:b1}
       B_1 ( y ) & = &  \frac{N-1}{2} \, \bt_1
                        - \frac{3}{2} \, \bar{g}_1 ( y ) \,,
   \makeatletter \if@double
   \hspace*{85mm}
   \fi\makeatother
       \\
       \lb{eq:b2}
       B_2 ( y ) & = & (N-1) \bt_2 + 9 \bar{g}_2 ( y )
                   + \frac{1}{16\pi^2}
                   \left( 3 \sqrt{3} \pi - N - 17
                          + 2 (N-1) \ln y \right)  \,, \\
       \lb{eq:b3}
       B_3 ( y ) & = & - \frac{N-1}{2} \bt_3
                       +  27 \left( \bar{g}_3 ( y ) +
                                 \frac{1}{32\pi^2 y^2} \right) \,.
\eqa
The  finite--size corrections $ \bar{g}_r $
are defined as follows:
\bea
      \lb{eq:gfunc}
      \bar{g}_r( m L) & = & L^{4-2r}
                          \left\{ \mlift \right. \frac{1}{L^4}
                           \sum_{ p\in \left( \frac{2\pi}{L} \ZZ \right)^4
                                }^{\hfill\prime\quad}
                            \left( p^2 + m^2 \right)^{-r}
                          - \int \!
                           \frac{d^4 \! p}{(2\pi)^4} \,
                            \left( p^2 + m^2 \right)^{-r}
                          \left. \mlift \right\} \non \\
             \mbox{}  & = & \left\{ \mlift \right.
                           \sum_{ n\in \ZZ ^4 }^{\hfill\prime\,}
                            \left( (2\pi n)^2 +  y^2 \right)^{-r}
                          - \int \!
                          \frac{d^4 \! z}{(2\pi)^4} \,
                            \left( z^2 + y^2 \right)^{-r}
                          \left. \mlift \right\}_{y=mL}
                           \,.
\eqa
\begin{figure}[ht]
\makeatletter
\if@double
\vspace*{-30mm}
\fi\makeatother
\vbox to 0mm{%
     \centerline{
         \setlength{\unitlength}{11mm}
         \begin{picture}(13,10)(0.00,0.00)
         \put(0.60,8.80){\makebox(0.0,0.0)[bl]{\large $\bar{g}_1(mL)$}}
         \put(6.85,0.35){\makebox(0.0,0.0)[bc]{\large $(mL)^2/4\pi$}}
         \put(7.00,3.10){\makebox(0.0,0.0)[bl]{:\quad continuum}}
         \put(7.00,3.70){\makebox(0.0,0.0)[bl]{:\quad $L=12$}}
         \put(5.90,3.09){\makebox(0.0,0.0)[bc]{\large{---}{---}{---}}}
         \put(6.00,3.09){\makebox(0.0,0.0)[bc]{\large{---}{---}{--}}}
         \put(5.90,3.69){\makebox(0.0,0.0)[bc]{\large -- -- -- --}}
         \end{picture}
      }\vss}
 \centerline{
 \epsfxsize=11cm
 \epsfbox{sfig1.ps}}
\makeatletter
\if@double
\vspace*{-13mm}
\fi\makeatother
\vbox to 0mm{%
     \centerline{
         \setlength{\unitlength}{11mm}
         \begin{picture}(13,10)(0.00,0.00)
         \put(0.60,8.80){\makebox(0.0,0.0)[bl]{\large $\bar{g}_2(mL)$}}
         \put(6.85,0.35){\makebox(0.0,0.0)[bc]{\large $(mL)^2/4\pi$}}
         \put(7.00,3.10){\makebox(0.0,0.0)[bl]{:\quad continuum}}
         \put(7.00,3.70){\makebox(0.0,0.0)[bl]{:\quad $L=12$}}
         \put(5.90,3.09){\makebox(0.0,0.0)[bc]{\large{---}{---}{---}}}
         \put(6.00,3.09){\makebox(0.0,0.0)[bc]{\large{---}{---}{--}}}
         \put(5.90,3.69){\makebox(0.0,0.0)[bc]{\large -- -- -- --}}
         \end{picture}
      }\vss}
 \centerline{
 \epsfxsize=11cm
 \epsfbox{sfig2.ps}
 }
\makeatletter
\if@double
\vspace*{-10mm}
\fi\makeatother
   \caption[The functions $\bar{g}_1 (mL)$ and  $\bar{g}_2 (mL)$]{
   The functions $\bar{g}_1 (mL)$ and  $\bar{g}_2 (mL)$ as defined
   in eq.~(\ref{eq:gfunc}) are shown as solid curves on the top
   and on the bottom, respectively.
   The dashed curves describe the modifications of these
   functions due to a finite lattice spacing in the case of the standard
   action, for a lattice size
   $L=12$.
   }
\lb{fig:latg}
\end{figure}
\makeatletter
\if@double\else
\clearpage
\fi\makeatother

These functions, which are discussed in more detail in appendix \ref{app:g},
are plotted in fig.~\ref{fig:latg} for $r\!=\!1,2$.
   For finite lattice spacing one has to replace $p^2$ by $\phat^2$ (in the
   case of the standard action) and $p$ is restricted to the Brillouin zone.
   Consequently, $\bar{g}_{r}$ is no longer a function of $mL$ alone and one
   gets different curves for different values of $L$. The continuum result
   is recovered as $m\!\to\! 0$, $L\!\to\! \infty$ with $mL$ fixed.
   However, as fig.~\ref{fig:latg} shows, even for $L=12$ the deviations are
   hardly visible. Hence lattice effects can be neglected and we shall use
   only the continuum form of $\bar{g}_{r}$.

The function
$\bar{g}_3$ shows a similar behaviour but is much smaller than
$\bar{g}_2$.
The so--called shape coefficients $ \beta_r $ are defined
in \cite{hale}. For $r=1,2,3$ they have the values
\be
         \beta_1 =  0.140461 \quad , \quad
         \beta_2 = -0.020305 \quad , \quad
         \beta_3 = -0.000482 \,.
\eq

Chiral perturbation theory expresses the leading behaviour of
$\chi$ as $L \!\to\! \infty$ in terms of a scale parameter
$\Lambda_\Sg$ \cite{gole} (see also \cite{neub}):
\be
               \chi
        =  \frac{N-1}{2}  \frac{Z^2}{\Sg^2}
         \left\{ \beta_2 + \frac{1}{8 \pi^2}
         \ln \left( \Lambda_\Sg L \right)   \right\}   \,.
\lb{eq:chia}
\eq
In order to compare with the perturbative formula (\ref{eq:chipert})
we neglect in the latter formula all terms which vanish for
$L \!\to\! \infty$. In this way we get
\be
          \chi                         =  \frac{1}{m_R^2}
     \left\{ 1 + \mlift \frac{g_R}{6}
     \left[ \frac{3 \pi \sqrt{3} - N -17}{16 \pi^2} +
     (N-1) \left( \slift \right. \beta_2
     + \frac{\ln \left( m_{R} L \right)}{8 \pi^2} \left. \slift \right)
     \right] \!+\! {\cal O} (g_R^2)
     \right\}.
\lb{eq:chib}
\eq
Equating (\ref{eq:chia}) and (\ref{eq:chib}) we find
\be
  \ln \Lambda_\Sg = \ln m_R  +
    \frac{48 \pi^2}{(N-1) g_R} +
    \frac{3 \pi \sqrt{3} - N -16}{2(N-1)} +     {\cal O} (g_R) \,,
\eq
where we have used
\be
 Z = 1 - \frac{1}{16 \pi^2} \frac{g_R}{6}  + {\cal O} (g_R^2) \,.
\eq
This formula is in complete agreement with a result applied in an earlier
investigation \cite{neuhaus}.

For the inverse propagator of the projected field in momentum space
we obtain at nonvanishing momentum $ p $
\bea
\lb{eq:invpro}
        \left[ G_{\pll} (p,-p) \right]^{-1} & = &
                      p^2 + m_R^2 \left[ \mlift \right.
                      1 + \frac{2 Z}{\Sg^2 L^2} B_1 ( m_{R} L)
               \\ \mbox{} & &
                      \hphantom{p^2 + m_R^2 \left[ 1 \mlift \right.}
                        \mbox{} - \frac{m_R^2 Z}{2 \Sg^2}
                          \left( B_2 ( m_{R} L) +
                          9 S_2 (m_R^2,p) + (N-1) S_2 (0,p)\right)
               \left. \mlift \right]
               \, , \non
\eqa
up to ${\cal O}( g_R^2 )$ terms,
where we have put
\be
      \lag \cph_{\pll} (x)  \cph_{\pll} (y) \rag_{j=0} =
           \frac{1}{L^4} \sum_{ p\in \left( \frac{2\pi}{L} \ZZ \right)^4}
                G_{\pll} ( p, -p) e^{ip(x-y)} \,.
\eq
Furthermore we have used the abbreviation
\renewcommand{\arraystretch}{0.8}
\be
          S_2 (m^2,p) = \frac{1}{L^4} \! \!
                      \sum_{ \scriptstyle \begin{array}{c} \scriptstyle
                      q\in \left( \frac{2\pi}{L} \ZZ \right)^4 \\
                      \scriptstyle q \neq 0, \, p \end{array}
                            } \!\!\!
                      \frac{1}{\left( q^2 + m^2 \right)
                      \left( (p-q)^2 + m^2 \right)} -
                      \frac{1}{L^4} \!\!
                      \sum_{ \scriptstyle \begin{array}{c} \scriptstyle
                      q\in \left( \frac{2\pi}{L} \ZZ \right)^4 \\
                      \scriptstyle q \neq 0 \end{array}
                            } \!\!\!
                      \frac{1}{\left( q^2 + m^2 \right)^2} \,.
\eq
\renewcommand{\arraystretch}{1.0}

\section{Methods for the determination of $\Sg$, $ Z $, and $ m_\sg $}
%
In this section we describe our procedure to determine
$\Sg$, $ Z $, and $ m_\sg $ from the data.
 Recall that all these quantities refer to the model at infinite
volume and vanishing external source.
As already mentioned in subsection \ref{sec:renorm} in our $1$--loop
calculation we can identify the renormalized $\sg$--mass $m_{R}$
with the physical mass $m_\sg$.

Lattice effects in the finite volume corrections $ \bar{g}_r (m L) $
turn out to be very small for the parameter values in our simulations
as is shown in fig.~\ref{fig:latg}.
Hence we shall neglect them and work with the continuum version,
so that the corrections are the same for both actions.
When applying formulas from the preceding section in which the field
normalization matters, one has, of course, to make sure that one uses the
correct normalization.
\subsection{Determination of $Z$}
\lb{sec:zfit}
The most effective way to calculate the Goldstone boson wave function
renormalization constant $ Z $ is to extract it from the correlation
function of the transverse field  $\lph^{\al}_{\perp} (x) $.
We measure the correlation function in momentum space
\bea
            G_{\perp} (p,-p) & = & \frac{1}{V} \frac{1}{N-1}
                \lag\sum_{\al} \tl{\lph}^{\al}_{\perp}(p)
                               \tl{\lph}^{\al}_{\perp}(-p)\rag
            2\kp (1+4\om) \,, \\
      \lph^{\al}_{\perp} (x) & = &
           \frac{1}{V} \sum_{ p\in \left( \frac{2\pi}{L} \ZZ_L \right)^4}
                 \tl{\lph}^{\al}_{\perp} (p)
                 e^{ipx}
\eqa
for some representative momenta and fit it according to
\be
            \left[ G_{\perp} (p,-p) \right]^{-1} =
                   Z^{-1} \left\{ m_{\pi}^2
                           + \frac{1}{1 + 4 \om}
                           \left( \phat^2 + \om \; \phati^2 \right)
                          \right\} \,.
\eq
Recall that $ \om \!=\! 0 $ corresponds to the standard nearest--neighbour
action.
Fits with two free parameters $ Z $, $ m_{\pi} $ led to very small values
for $ m_{\pi} $ on all of our volumes indicating
that $ G_{\perp} (p,-p)  $ is indeed dominated by the Goldstone bosons.
As our final  values  for $ Z $ we take the results from a 1--parameter
fit with $ m_{\pi} $ fixed to zero.
They agree with the values of the 2--parameter fit within statistical
errors.
We have not seen any finite--size dependence of $Z$.
%
\begin{figure}[ht]
\vbox to 0mm{%
  \centerline{
      \setlength{\unitlength}{11mm}
      \begin{picture}(13,10)(0.00,0.00)
      \put(6.72,0.35){\makebox(0.0,0.0)[bc]{\large $1/L^{\,2}$}}
      \put(0.48,8.55){\makebox(0.0,0.0)[bl]{\large $\tlag\Mc\trag_{j=0}$}}
      \put(3.2,7.25){\makebox(0.0,0.0)[bl]{$\kp$}}
      \put(3.2,6.65){\makebox(0.0,0.0)[bl]{$\Sg$}}
      \put(3.2,6.05){\makebox(0.0,0.0)[bl]{$Z$}}
      \put(3.2,5.45){\makebox(0.0,0.0)[bl]{$m_\sg$}}
      \put(3.8,7.25){\makebox(0.0,0.0)[bl]{$= 0.308$}}
      \put(3.8,6.65){\makebox(0.0,0.0)[bl]{$= 0.12827(6)$}}
      \put(3.8,6.05){\makebox(0.0,0.0)[bl]{$= 0.972$}}
      \put(3.8,5.45){\makebox(0.0,0.0)[bl]{$= 0.3223$}}
      \end{picture}
      }\vss}
 \centerline{
 \epsfxsize=11cm
 \epsfbox{sfig3.ps}
 }
   \caption[The mean field as function of
         $1/L^2$ at $\kp \!=\! 0.308$.]{The
         mean field $\tlag \Mc \trag_{j=0}$ as function of
         $1/L^2$ at $\kp \!=\! 0.308$.
         The solid curve represents our fit.
         Its intercept in the infinite volume
         limit gives $\Sg$, shown as a dotted line.
         The points scattered around this line are obtained by subtracting
         the finite--volume corrections from the data.
         }
\lb{fig:sigfs}
\end{figure}
\subsection{Determination of $\Sg$}
\lb{sec:sfit}
  $ \Sg $ has been determined in two ways, either from the expectation
value of $ \Mc $ at $ j=0 $ or from the expectation value of $
\Mc_N $ for nonvanishing $ j $. These expectation values are determined
from the double distribution of the magnetization and the energy, see eqs.\
(\ref{eq:distproj}) and (\ref{eq:distabs}).

The connection with $\Sg$ is established through the perturbative expression
for the minimum $\Mc_0$ of the effective potential
(see eq.~(\ref{eq:mineff})),
\be
         \Mc_0 \! = \! \Sg
                   \left( 1 + B_1 ( m_{\sg} L) \frac{Z}{\Sg^2 L^2}
                   \right) \,,
\eq
which will be used
in eqs.~(\ref{eq:mnj}), (\ref{eq:mnk}),
\bea
    \lb{eq:sig}
     \tlag \Mc \trag_{j=0} & = & \Mc_0
                   \left( \mlift \right.
                   1 + \frac{1}{V \Mc_0^2}
                       \left( (N - 1) \chi \tlift + 3a_2 \Mc_0 \right)
                   \left. \mlift \right) \,, \\
     \tlag \Mc_N \trag_{j} & = &
                   \frac{1}{jV} \left( \frac{u Y_N'(u)}{Y_N(u)}
                         \right)
                                + j \chi \,,
    \lb{eq:sigj} \\
    \mbox{} & & u = V j \Mc_0
                    \left\{ 1 + \slift \right.
                    \frac{1}{V \Mc_0^2} \,
                    \left( \frac{N-1}{2} \chi
                   + \tlift 3 a_2 \Mc_0 \right) \left. \slift \right\}
                   \,, \non
\eqa
together with the  perturbative results
(\ref{eq:chipert}),
(\ref{eq:a2m}) for $\chi = a_1^2$ and $3a_2\Mc_0$, respectively.
Remember that within our accuracy $\chi$ does not depend on $j$.

In order to determine $ \Sg $ from either of those equations
    (\ref{eq:sig}),  (\ref{eq:sigj}) we need $ Z $
and $ m_\sg $. We take $ Z $ from the  propagator fit discussed in subsection
\ref{sec:zfit}.
For $ m_\sg $ we can either insert values obtained as described in
 subsection \ref{sec:mfit} or we can leave it as a fit parameter.
The corresponding differences in $ \Sg $ are of the same order of magnitude
as the statistical errors.
In any case, the results are not very sensitive to the value of $ m_\sg $
since it enters only through the $1$--loop corrections.

As our final method for the determination of $\Sg$ we have
adopted the approach using eq.~(\ref{eq:sig}),
since in this case we do not need an analytic continuation in $j$.
\begin{figure}[ht]
\vbox to 0mm{%
  \centerline{
      \setlength{\unitlength}{11mm}
      \begin{picture}(13,10)(0.00,0.00)
      \put(6.72,0.45){\makebox(0.0,0.0)[bc]{\large $1/L^{\,2}$}}
      \put(1.20,8.55){\makebox(0.0,0.0)[bl]{\Large $\chi$}}
      \put(7.4,7.25){\makebox(0.0,0.0)[bl]{$\kp$}}
      \put(7.4,6.65){\makebox(0.0,0.0)[bl]{$\Sg$}}
      \put(7.4,6.05){\makebox(0.0,0.0)[bl]{$Z$}}
      \put(7.4,5.45){\makebox(0.0,0.0)[bl]{$m_\sg$}}
      \put(8.0,7.25){\makebox(0.0,0.0)[bl]{$= 0.308$}}
      \put(8.0,6.65){\makebox(0.0,0.0)[bl]{$= 0.12827$}}
      \put(8.0,6.05){\makebox(0.0,0.0)[bl]{$= 0.972$}}
      \put(8.0,5.45){\makebox(0.0,0.0)[bl]{$= 0.3223(12)$}}
      \end{picture}
      }\vss}
 \centerline{
 \epsfxsize=11cm
 \epsfbox{sfig4.ps}
 }
   \caption[The susceptibility $ \chi $ as function  of $1/L^2$ at
         $\kp \!=\! 0.308$.]{The susceptibility $ \chi $ as function  of
         $1/L^2$ at $\kp \!=\! 0.308$.
         The curve represents our fit.
         The $ 1/L^2 $ contribution dominates, i.e.\  for the simulated
         lattice sizes the susceptibility is not governed  by a logarithmic
         behaviour as it is the case in chiral perturbation theory.}
\lb{fig:masfs}
\end{figure}

\subsection{Determination of $m_\sg$}
\lb{sec:mfit}
We have employed two methods in order to calculate $ m_\sg $.
The first method uses the propagator of the projected field
$ \cph_{\pll}  (x) $. We fit the propagator in momentum space with
a formula analogous to the case of the transverse field,
\be
            \left[ G_{\pll} (p,-p) \right]^{-1} =
                   Z^{-1}_{\sg} \left\{ m_{\sg}^2 (L)
                           + \frac{1}{ 1 + 4 \om }
                           \left( \phat^2 + \om \; \phati^2 \right)
                          \right\} \,.
    \lb{eq:msigp}
\eq
The $L$--dependent mass $ m_\sg (L) $ is extrapolated to infinite volume
by means of the formula
\be
          Z^{-1}_{\sg} m_{\sg}^2 (L) = m_{\sg}^2
                \left( 1 + \frac{2 Z}{\Sg^2 L^2} B_1 ( m_{\sg} L)
                         - \frac{m_\sg^2 Z}{2 \Sg^2} B_2 ( m_{\sg} L)
                \right) \,,
    \lb{eq:msigl}
\eq
which is suggested by eq.~(\ref{eq:invpro}) in the limit of
small momenta $p$.

A more accurate procedure  to determine $ m_\sg $ uses the
susceptibility $ \chi $.
We fit the numerical results for $ \chi $ according to
(compare eq.~(\ref{eq:chipert})):
\be
 \lb{eq:chifit}
         \chi = \frac{1}{m_\sg^2} \,
                    \left( 1 - \frac{2 Z}{\Sg^2 L^2} B_1 ( m_\sg L)
                          \right) \,
                     \left( 1 + \frac{m_\sg^2 Z}{2 \Sg^2} B_2 ( m_\sg L)
                         \right) \,. \non
\eq
To this end we need $ Z $ and $ \Sg $.
As above, we take $ Z $ from the  propagator fit discussed in
subsection \ref{sec:zfit}.
For $ \Sg $ we can either insert values obtained by
the methods described in subsection \ref{sec:sfit} or
we can leave it as a fit parameter.
The corresponding differences in $ m_\sg $ are of the same order of
magnitude as the statistical errors.

The procedure to extract $ \Sg $ from the magnetization and $ m_\sg $
from the susceptibility is performed iteratively until the
results stabilize.
Examples of the final fits are shown  in figures \ref{fig:sigfs} and
\ref{fig:masfs} for the standard action at $ \kp \! =\! 0.308 $.
\subsection{Systematic errors due to higher loop corrections}
%
\lb{sec:loop}
In all the above perturbative formulas we have restricted ourselves
to 1-loop order. Estimates on the influence of the neglected higher
loop contributions may be obtained from the following observations.

In the case of the determination of $\Sigma$ we can compare our results
based on eq.~(\ref{eq:sig}) with the numbers extracted from
eq.~(\ref{eq:sigj}).
The 2-loop contributions should be different, yet the
values for $\Sigma$
do not differ by more than 2 times the statistical error.  We
take this as an indication that the use of 2-loop formulas would
not alter our results for $\Sigma$ by more than about $0.2$ percent.

Turning to our calculation of $m_\sigma$ we have to note the specific
form~(\ref{eq:chifit}) in which the perturbative formula for $\chi$
is used to fit the data.
Application of the formula
\be
\chi = \frac{1}{m_\sigma^2} \left(1 - \frac{2Z}{\Sigma^2 L^2}
       B_1 (m_\sigma L) + \frac{m_\sigma^2 Z}{2 \Sigma^2}
       B_2 (m_\sigma L) \right) \quad ,
\eq
which is equivalent to (\ref{eq:chifit}) in 1-loop order, leads to a less
satisfactory description of the $L$-dependence. Nevertheless, the
results for $m_\sigma$ would be changed only by about $1.5$ percent.
Furthermore, the mass values extracted from the propagator of the
projected field according to eqs.(\ref{eq:msigp})
and (\ref{eq:msigl}) agree with the more
accurate results from the susceptibility rather well: The differences
range from less than one up to a few standard deviations, the
agreement being better for the larger $\kappa$-values. Therefore, we
expect that the systematic uncertainties of $m_\sigma$ do not exceed
the statistical errors significantly.

\section{Algorithm and numerical results}
%
In order to generate our configurations
we applied the reflection cluster algorithm  for vanishing external source
\cite{wolff} and implemented the additional possibility of handling
kinetic terms with  couplings over two lattice spacings (see
eq.~(\ref{eq:lact})).

After choosing randomly a direction $\itvec{r}$ in the space of the
internal $O(4)$--symmetry, we connect two points $ x $, $ x+\hmu $ by a bond
with probability
\be
    p_{(x,x+\hmu)} = \mbox{max} \left\{ \slift \right.
                   0, 1- \exp
                   \left( - 4 \kp r^\al \lph^\al  (x)
                                  r^\bt \lph^\bt (x+\hmu) \right)
                                \left. \slift \right\} \,.
\eq
The points $ x $ and $ x + 2\hmu $ are linked with probability
\be
    p_{(x,x+2 \hmu)} = \mbox{max} \left\{ \slift \right.
                   0, 1- \exp
                   \left( - 4 \kp \om r^\al \lph^\al  (x)
                                  r^\bt \lph^\bt (x+2 \hmu) \right)
                                \left. \slift \right\} \,.
\eq
In order to find the corresponding clusters we used
a standard tree search algorithm, which could be efficiently
vectorized.
The clusters were flipped with probability $ 1/2 $,
where a flip means a reflection of the field variables
with respect to the hyperplane orthogonal to $ \itvec{r} $.

The simulation is performed on hypercubic lattices
of volumes $ V\! =\! L^4 $ ranging from $L\! =\! 12$ to $L\! =\! 20$.
The accumulated statistics on each lattice and $\kp$--value
was about $ 20\,000 - 150\,000$ measurements,
where successive measurements are separated by 4 multi--cluster
configurations. The statistical errors have been determined by a blocking
procedure: We have divided the data into blocks of $5\,000 - 6\,000$ data
points and treated the averages (and distributions)
on each block as independent measurements.
%
\begin{figure}[ht]
\vbox to 0mm{%
  \centerline{
      \setlength{\unitlength}{11mm}
      \begin{picture}(13,10)(0.00,0.00)
      \put(6.90,0.60){\makebox(0.0,0.0)[bc]{\large $E$}}
      \end{picture}
      }\vss}
  \centerline{
  \epsfxsize=11cm
  \epsfbox{sfig5.ps}
  }
   \caption[Distribution of the kinetic energy
            for $L \! =\! 16$ in the case of the standard
            action]{Distribution of the kinetic energy
            (see eq.~(\ref{eq:o}))
            for $L \! =\! 16$ in the case of the standard action
            }
\lb{fig:distri}
\end{figure}

We have measured a discretized version of the double distribution
 $ \tl{Z}_{ \kp_0 } ( \Ml , \El ) $  of the kinetic
energy and the magnetization for a set of $\kp_0$--values.
We have chosen $\kp_0 \! =\!  0.306$, $0.308$, $0.310$ and
$\kp_0 \! =\! 0.338$, $0.340$, $0.345$, $0.350$ in the case of the
standard and the Symanzik improved action, respectively.
We analytically continue our results to $ \kp $-values in the neighbourhood
of $ \kp_0 $ as well as to nonzero values of the external source $ J $ as
explained in sec.\ \ref{sec:distri} (so--called single--histogram method).

As shown in fig.~\ref{fig:distri} for the standard action at
$L\!=\! 16$ the individual distributions for the different
$\kp_0$--values do not overlap very much.
Hence the multi--histogram method would not lead to significant
improvements.

In tables \ref{tb:stand} and \ref{tb:impro} we summarize our results
for $ Z $, $ m_\sg $, $ \Sg $ obtained by the methods described in the
previous sections in the case of the standard and the improved action,
respectively. The errors quoted in these tables are purely statistical as
returned by the fitting routine. Note that the improved action
contains a term with a small antiferromagnetic coupling, thus
a reduction of the efficiency of the cluster algorithm is possible.
However, comparing typical errors of $\Sg$ and $m_\sg$
in tables \ref{tb:stand} and \ref{tb:impro} we note
that the cluster algorithm performs as well in the case of the improved
action as it does for the standard action.

\makeatletter \if@double
\renewcommand{\arraystretch}{0.7}
\else
\renewcommand{\arraystretch}{1.3}
\fi\makeatother
\begin{table}[ht]
   \bc
   \begin{tabular}{||c|c|c|c||}
   \hline \hline \mlift
      \hphantom{xxx}       $\kp$     \hphantom{xxx}   &
      \hphantom{xxxxx}     $\Sg$     \hphantom{xxxxx} &
      \hphantom{xxxx}     $m_\sg$    \hphantom{xxxx}  &
      \hphantom{xxx}       $Z$       \hphantom{xxx}  \\ \hline
   0.3056   &  0.08030(16)    & 0.1742(7)\zr  &   \ulift      \\
   0.3058   &  0.08545(14)    & 0.1918(7)\zr  &               \\
   0.3060   &  0.09045(10)    & 0.2072(7)\zr  &   0.977(3)    \\
   0.3062   &  0.09503(11)    & 0.2210(7)\zr  &               \\
   0.3064   &  0.09929(12)    & 0.2342(8)\zr  &               \\
   0.3066   &  0.10334(13)    & 0.2472(10)    &               \\
   0.3068   &  0.10722(14)    & 0.2600(11)    &               \\
   0.3070   &  0.11090(16)    & 0.2724(14)    &               \\
   0.3072   &  0.11505(19)    & 0.2847(23)    &               \\
   0.3074   &  0.11851(14)    & 0.2929(20)    &               \\
   0.3076   &  0.12187(11)    & 0.3024(17)    &               \\
   0.3078   &  0.12512(8)\zr  & 0.3124(14)    &               \\
   0.3080   &  0.12827(6)\zr  & 0.3223(12)    &   0.972(3)    \\
   0.3082   &  0.13130(6)\zr  & 0.3322(11)    &               \\
   0.3084   &  0.13424(8)\zr  & 0.3420(11)    &               \\
   0.3086   &  0.13710(8)\zr  & 0.3517(12)    &               \\
   0.3088   &  0.13987(9)\zr  & 0.3614(14)    &               \\
   0.3090   &  0.14268(19)    & 0.3727(29)    &               \\
   0.3094   &  0.14786(14)    & 0.3833(13)    &               \\
   0.3096   &  0.15051(12)    & 0.3907(14)    &               \\
   0.3098   &  0.15305(10)    & 0.3977(18)    &               \\
   0.3100   &  0.15549(10)    & 0.4073(17)    &   0.964(7)    \dlift \\
   \hline \hline
   \end{tabular}
   \ec
   \caption[Results for the standard action]{Results
            for the standard action.
            The quoted errors are purely statistical.}
   \lb{tb:stand}
\end{table}
\renewcommand{\arraystretch}{1.0}
%
\makeatletter \if@double
\renewcommand{\arraystretch}{0.7}
\else
\renewcommand{\arraystretch}{1.3}
\fi\makeatother
\begin{table}[ht]
   \bc
   \begin{tabular}{||c|c|c|c||}
   \hline \hline \mlift
      \hphantom{xxx}       $\kp$     \hphantom{xxx}   &
      \hphantom{xxxxx}     $\Sg$     \hphantom{xxxxx} &
      \hphantom{xxxx}     $m_\sg$    \hphantom{xxxx}  &
      \hphantom{xxx}       $Z$       \hphantom{xxx}  \\ \hline
   0.33725  &  0.06980(43)    & 0.1696(53)      &   \ulift      \\
   0.33750  &  0.07668(43)    & 0.1813(31)      &               \\
   0.33775  &  0.08333(38)    & 0.1985(25)      &               \\
   0.33800  &  0.08887(31)    & 0.2184(30)      &   0.971(4)    \\
   0.33825  &  0.09363(32)    & 0.2322(27)      &               \\
   0.33850  &  0.09805(37)    & 0.2496(31)      &               \\
   0.33875  &  0.10212(40)    & 0.2739(37)      &               \\
   0.33925  &  0.10927(27)    & 0.2960(47)      &               \\
   0.33950  &  0.11295(19)    & 0.3065(38)      &               \\
   0.33975  &  0.11647(15)    & 0.3179(30)      &               \\
   0.34000  &  0.11983(12)    & 0.3299(25)      &   0.968(3)    \\
   0.34025  &  0.12311(10)    & 0.3413(24)      &               \\
   0.34050  &  0.12636(9)\zr  & 0.3523(25)      &               \\
   0.34075  &  0.12941(12)    & 0.3656(29)      &               \\
   0.34450  &  0.16644(12)    & 0.5186(35)      &               \\
   0.34475  &  0.16831(7)\zr  & 0.5231(31)      &               \\
   0.34500  &  0.17037(7)\zr  & 0.5291(27)      &   0.963(3)    \\
   0.34525  &  0.17246(9)\zr  & 0.5364(25)      &               \\
   0.34550  &  0.17446(9)\zr  & 0.5449(24)      &               \\
   0.34575  &  0.17638(10)    & 0.5529(26)      &               \\
   0.34600  &  0.17827(12)    & 0.5605(29)      &               \\
   0.34625  &  0.18015(15)    & 0.5694(33)      &               \\
   0.34950  &  0.20286(8)\zr  & 0.6611(52)      &               \\
   0.34975  &  0.20450(7)\zr  & 0.6655(45)      &               \\
   0.35000  &  0.20612(6)\zr  & 0.6724(38)      &   0.959(3)    \\
   0.35025  &  0.20769(6)\zr  & 0.6819(37)      &               \\
   0.35050  &  0.20924(7)\zr  & 0.6895(40)      &               \\
   0.35075  &  0.21076(8)\zr  & 0.6968(44)      &   \dlift      \\
   \hline \hline
   \end{tabular}
   \ec
   \caption[Results for the improved action]{Results for the
            improved action.
            The given errors are purely statistical.}
   \lb{tb:impro}
\end{table}
\renewcommand{\arraystretch}{1.0}
%
\begin{figure}[ht]
\vbox to 0mm{%
  \centerline{
      \setlength{\unitlength}{11mm}
      \begin{picture}(13,10)(0.00,0.00)
      \put(7.20,0.40){\makebox(0.0,0.0)[bc]{\Large $\kp$
      \large $^{\rm (improved)}$}}
      \put(7.20,9.15){\makebox(0.0,0.0)[bc]{\Large $\kp$
      \large $^{\rm (standard)}$}}
      \put(0.85,8.70){\makebox(0.0,0.0)[bl]{\Large $\Sg^{\,2}$}}
      \put(3.00,7.40){\makebox(0.0,0.0)[bl]{$\kp_c$}}
      \put(3.60,7.40){\makebox(0.0,0.0)[bl]{$= 0.30423(1)$}}
      \put(6.00,3.40){\makebox(0.0,0.0)[bl]{$\kp_c$}}
      \put(6.60,3.40){\makebox(0.0,0.0)[bl]{$= 0.33592(1)$}}
      \end{picture}
      }\vss}
 \centerline{
 \epsfxsize=11cm
 \epsfbox{sfig6.ps}
 }
   \caption[The square of the field expectation value $\Sg$
      as function of $\kp$ for the standard and the Symanzik improved
      action, respectively.]{ The square of the field expectation value
      $\Sg$ as function of $\kp$ for the standard (left curve, upper
      $\kp$--scale) and the Symanzik improved  action (right curve, lower
      $\kp$--scale), respectively.
      The solid curves are the scaling law fits.
      In the case of the standard action we can compare with the results of
      L\"uscher and Weisz \cite{luwe}:
      Our $\Sg^2$--values are situated at the lower border of their
      results (dotted curve).}
\lb{fig:sigscal}
\end{figure}

\makeatletter
\if@double\else
\pagebreak[2]
\fi\makeatother
\section{Scaling behaviour}
Our results presented in the preceding section will now be used to discuss
the scaling behaviour of the model.

{}From perturbation theory one expects mean field scaling laws
with logarithmic corrections.
Approaching the critical point in the broken phase
one finds in the $O(4)$--model
\bea
           \Sg^2   & = & C_\Sg \tau \, |\ln (\tau)|^{1-\al}
            \lb{eq:sigscal} \,, \\
           m_\sg^2 & = & C_m \tau \, |\ln (\tau)|^{-\al} \, ,
            \lb{eq:masscal} \\
                Z  & = & C_Z
\eqa
in terms of $\tau \! =\!  \kp/\kp_{c} -1$
with $C_\Sg, C_m, C_Z = const$.
The exponent $ \al $ is predicted to be $\al \!=\! 1/2$
\cite{blz}.

According to eq.~(\ref{eq:sigscal}) we performed
a $3$--parameter fit to the $ \Sg^2 $--data with
$ C_\Sg $, $\kp_c$ and $ \al $ as free parameters
and used the results as input
for a $1$--parameter fit to the $ m_\sg $--data in order to extract
$ C_m $.
By this procedure we find
$\kp_{c} \! =\! 0.30423(1)$, $ \al \!=\! 0.63(1) $,
$C_\Sg \!=\! 0.877(6)$, $C_m  \!=\! 4.60(1)$
for the standard action and
$\kp_{c} \! =\! 0.33592(1)$, $ \al \!=\! 0.54(1) $,
$C_\Sg \!=\! 0.770(2)$, $C_m  \!=\! 4.49(1)$
for the Symanzik improved action.
The errors
might be underestimated since we have neglected correlations
between the data points.
Varying the number of data points used in the fit changes $\kp_c$ only
within the quoted errors while the changes of the exponent $\al$ are about
five times the statistical errors.
%
\begin{figure}[ht]
\vbox to 0mm{%
  \centerline{
      \setlength{\unitlength}{11mm}
      \begin{picture}(13,10)(0.00,0.00)
      \put(7.20,0.40){\makebox(0.0,0.0)[bc]{\Large $\kp$
      \large $^{\rm (improved)}$}}
      \put(7.20,9.15){\makebox(0.0,0.0)[bc]{\Large $\kp$
      \large $^{\rm (standard)}$}}
      \put(0.80,8.70){\makebox(0.0,0.0)[bl]{\Large $m_\sg^2$}}
      \end{picture}
      }\vss}
 \centerline{
 \epsfxsize=11cm
 \epsfbox{sfig7.ps}
 }
   \caption[The square the $\sg$--mass $m_\sg$ as function of $\kp$ for the
      standard and the Symanzik improved action, respectively.]{The
      square the $\sg$--mass $m_\sg$ as function of $\kp$ for the
      standard (left curve, upper $\kp$--scale) and the Symanzik improved
      action (right curve, lower $\kp$--scale), respectively.
      The solid curves are the scaling law fits.
      In the case of the standard action we can compare with the results of
      L\"uscher and Weisz \cite{luwe}:
      Our $\sg$--masses lie between their $2$--loop (dotted curve) and
      $3$--loop (dashed curve) predictions.}
\lb{fig:masscal}
\end{figure}

Thus, in the case of the improved action
we obtain satisfactory agreement with the perturbative predictions.
However, in the case of the standard action
the fit--result for $ \al $ indicates the need for a more sophisticated
treatment if consistency with perturbation theory is to be established
(leaving aside the more exotic possibility of nonperturbative effects).
For the standard action we therefore
integrate the renormalization group
equations using the perturbative $3$--loop
expansions of ref.~\cite{luwe}. Note that these contain scaling violations
at the $1$--loop level.

This requires the calculation of the renormalization constant
$Z_R^{\cal O}$ associated with the composite field ${\cal O} (x)$
defined in eq.~(\ref{eq:o}), which was not measured directly in our
simulations.
Choosing two data points $(\Sg$, $m_\sg$, $Z$$)_{1,2}$
-- corresponding to ($v_R$, $m_R$, $Z_R$) in \cite{luwe} --
we determine $(Z_R^{\cal O} )_1$ such that the integration
of the renormalization group equations starting from point $1$
reproduces point $2$ as closely as possible.
For point $ 1 $ we took the results at $ \kp \!=\! 0.306 $, for point $ 2 $
$ \kp \!=\! 0.308 $.
 The agreement of the data between these points with the results from the
integration is as good as with the scaling law fits.
 This indicates that the deviation of the fit parameter $ \al $
from $ 1/2 $ in the case of the standard action is due to nonnegligible
higher loop contributions or scaling violations
in the renormalization group functions
invalidating the scaling laws
(\ref{eq:sigscal}), (\ref{eq:masscal}) for the larger $\kp$--values.
For the improved action
on the other hand such contaminations seem to be suppressed in the whole
range of $ \kp $--values studied.

\makeatletter \if@double
\renewcommand{\arraystretch}{0.8}
\else
\renewcommand{\arraystretch}{1.3}
\fi\makeatother
\begin{table}[ht]
   \bc
   \begin{tabular}{||c|c|c|c|c||}
   \hline \hline \mlift
      \hphantom{xxx}       $\kp$     \hphantom{xxx}   &
      \hphantom{xxxx}     $\Sg$     \hphantom{xxxx} &
      \hphantom{xxx}     $m_\sg$    \hphantom{xxx}  &
      \hphantom{xxx}     $Z$        \hphantom{xxx}  &
      \hphantom{xxxxxxxxxx}                  \\ \hline
  0.3060\phn{(2)}
            & 0.09045(10)       & 0.2072(7)\zr     & 0.977(3) &
                                                         this work\ulift\\
  0.3059(2) & 0.097(3)\phn{540} & 0.210\phn{0(70)} & 0.972(6) & 2--loop \\
  0.3060(2) & 0.096(3)\phn{540} & 0.210\phn{0(70)} & 0.972(6) & 3--loop \\
  \hline
  0.3080\phn{(3)}
            & 0.12827(6)\zr     & 0.3223(12)       & 0.972(3) & this work \\
  0.3079(3) & 0.139(4)\phn{020} & 0.320\phn{2(12)} & 0.973(6) & 2--loop\\
  0.3081(3) & 0.137(4)\phn{020} & 0.320\phn{2(12)} & 0.973(6) & 3--loop \\
  \hline
  0.3100\phn{(5)}
            & 0.15549(10)       & 0.4073(17)       & 0.964(7) & this work \\
  0.3100(5) & 0.170(6)\phn{290} & 0.410\phn{2(17)} & 0.974(7) & 2--loop \\
  0.3101(5) & 0.164(6)\phn{290} & 0.400\phn{2(17)} & 0.974(7) & 3--loop
  \dlift \\
   \hline \hline
   \end{tabular}
   \ec
   \caption[Comparison between the analytic results of L\"uscher and Weisz
      and the data of our numerical simulation]{Comparison between  the
      analytic results of L\"uscher and Weisz (second and
      third line) and the data of our numerical simulation (first line).
      The $2$--loop and $3$--loop values were calculated by us using the
      methods leading to table~8 in ref.~\cite{luwe}.}
   \lb{tb:comp}
\end{table}
\renewcommand{\arraystretch}{1.0}
%
In figures \ref{fig:sigscal} and \ref{fig:masscal}
the squares of the field expectation value $\Sg$ and the
$\sg$--mass $m_\sg$ are shown as functions of $\kp$ for the
standard (left curves, upper $\kp$--scale) and the Symanzik improved action
(right curves, lower $\kp$--scale), respectively.
The solid curves are scaling law fits.
In the case of the standard action we can compare with the results of
L\"uscher and Weisz \cite{luwe}:
Our $\Sg^2$--values are situated at the lower border of their
results (dotted curve),
our $\sg$--masses lie between their $2$--loop (dotted curve) and $3$--loop
(dashed curve) predictions.

In table \ref{tb:comp} we summarize the comparison
for some significant $\kp$--values.
The large errors of the analytical $\Sg$--results are due to
the uncertainties from the high--temperature expansion and
from the perturbation expansion of the $\bt$--function
arising in the $g_R$ determination.
Our numerical determination of $\Sg$ appears to be more precise.

\section{Discussion of the upper bound on the Higgs mass}
In fig.~\ref{fig:bound} we show our results for the quantity $m_\sg/F$ as
a function of the correlation length $1/m_\sg$ of the $\sg$--particle,
compared with a scaling behaviour of the form expected from the perturbative
$2$--loop $\beta$--function:
\be
         \frac{1}{m_\sg} =
               \frac{1}{C} \;
               \left( \frac{\ds F^2}{\ds 3 c_1 m_\sg^2} \right)^{c_2}
               \; \exp \left( \frac{\ds F^2}{\ds 3 c_1 m_\sg^2} \right) \,.
\lb{eq:betascal}
\eq
In the case of the O(4)--model perturbation theory leads to
$c_1 \!=\! \bt_1 \!=\! 1/(4\pi^{2})$, $c_2 \!=\! - \bt_2/\bt_1^{\,2} \!=\!
13/24$.

Since our numerical results cover only a small interval of
$m_{\sg}/F$--values, it is not feasible to determine $c_2$ from a fit
with the ansatz (\ref{eq:betascal}). However, $C$ and $c_1$ can be estimated
by a fit with $c_2\!=\!0$ corresponding to a $1$--loop $\bt$--function. In
order to get an impression of the systematical errors, we have performed a
second fit with $c_2 \!=\! 13/24$ as predicted by $2$--loop
perturbation theory.
The results are shown in table \ref{tb:beta}.
The agreement of the fit results for $\bt_1$ with the expectation from
perturbation theory is satisfactory. For both actions the perturbative value
turns out to lie between the numbers derived from the two fits.
(Again the quoted errors may be too small, because correlations have been
ignored.) In the case of the standard action the result for $\ln C$,
however, deviates from the prediction by L\"uscher and Weisz \cite{luwe}
almost by a factor of two.

%
\begin{figure}[ht]
\vbox to 0mm{%
  \centerline{
      \setlength{\unitlength}{11mm}
      \begin{picture}(13,10)(0.00,0.00)
      \put(6.88,0.20){\makebox(0.0,0.0)[bc]{\large $1/m_\sg$}}
      \put(0.80,8.30){\makebox(0.0,0.0)[bl]{\large
      $\frac{\ds \tlift m_{\sg}}{\ds \tlift F}$}}
      \end{picture}
      }\vss}
 \centerline{
 \epsfxsize=11cm
 \epsfbox{sfig8.ps}
 }
   \caption[The ratio $m_\sg / F$
      for the Symanzik improved action and
      for the standard action, which can be compared with
      the results of L\"uscher and Weisz (table~8)]{The data points of the
      ratio $m_\sg / F$ for the Symanzik improved action lie above the
      values for the standard action, which can be compared with
      the results in table~8 of L\"uscher and Weisz \cite{luwe} (crosses).
      According to the discussion of the errors in that reference we
      plot their data with doubled errors.
      Solid curves are fits with respect to a scaling law of the form
      predicted by perturbation theory.}
\lb{fig:bound}
\end{figure}
\makeatletter \if@double
\renewcommand{\arraystretch}{0.8}
\else
\renewcommand{\arraystretch}{1.3}
\fi\makeatother
\begin{table}[ht]
   \bc
   \begin{tabular}{||c||c|c||c|c||c||}
   \hline \hline
   \hphantom{xxxxx} &
   \multicolumn{2}{c||}{\mlift \bf
            Improved action} &
   \multicolumn{3}{||c||}{\mlift \bf
            Standard action} \\
   \hline \mlift
      \mbox{} &
         $c_2=0$      &
         $c_2=13/24$  &
         $c_2=0$      &
         $c_2=13/24$  &
         refs.~\cite{blz,luwe}\\ \hline \hline
  $\bt_1$   & 0.0212(1)    & 0.0292(2)
            & 0.0236(1)    & 0.0303(1)
            & 0.02533\phn{(1)} \ulift \\ \hline
  $\ln C$   & 1.066(7)\zr  & 0.713(10)
            & 1.174(9)\zr  & 0.982(11)
            & 1.9(1)\phn{2533} \dlift \\ \hline \hline
   \end{tabular}
   \ec
   \caption[Values of $C$ and the $\bt_1$--coefficient of the
      $\bt$--function.]{Values of $C$ and the $\bt_1$--coefficient of the
      $\bt$--function.
      The first column shows the results corresponding to
      the $1$--loop formula. In the second column we have taken into account
      the $2$--loop corrections with $c_2$ fixed to $13/24$.
      In the case of the standard action we can compare our $\ln C$--values
      with the result of L\"uscher and Weisz \cite{luwe}.}
   \lb{tb:beta}
\end{table}
\renewcommand{\arraystretch}{1.0}

The data of fig.~\ref{fig:bound} can be used to determine the upper bound on
the mass of the $\sg$--particle from the requirement that $m_\sg$ be not
too close to the cut--off. It is obvious that a condition of this
kind has to be imposed for the model to make sense as an effective
field theory, because a particle mass above the cut--off is physically
unacceptable. However, to make this requirement (and hence the
definition of the upper bound) more precise different procedures
have been proposed. For instance, one might choose a (sufficiently
small) fixed value for $m_\sg$ (e.g.\ $m_\sg \!=\! 0.5$) where cut--off
effects are assumed to have become negligible. But this criterion
does not take into account that the amount of scaling violations
depends on the regularization, i.e.\ on the lattice action.

It seems therefore to be more sensible to replace the limit on
$m_\sg$ by a prescribed bound on scaling violations. For example,
one can study the scattering of the Goldstone bosons. Since it is very
hard to extract scattering amplitudes from numerical simulations,
one has to resort to perturbative calculations. This should lead
to reliable results after all, because perturbation theory is
known to describe the model successfully in the (approximate)
scaling region (which is all we are interested in at the moment)
and high precision is not required in this somewhat qualitative
discussion. We have therefore followed L\"uscher and Weisz \cite{luwe}
and calculated the differential cross section for
Goldstone boson scattering in the center--of--mass frame at tree
level of perturbation theory. From the action (\ref{eq:actcont}) with
$j=0$ we obtain for the scattering of two Goldstone bosons
carrying the same $O(N)$--index into any pair of two other Goldstone
bosons
\bea
   \frac{d \sg}{d \Omega} & = & \frac{|\vp^\prime|^2}{16 \pi ^2}
    \frac{1}{|E^\prime (\vp) E^\prime (\vp^\prime)|}
     \left[ \frac{1 + 4 \om}{2 \sinh(W/2) + 4 \om \sinh W } \right] ^4
     \nonumber \\ {} &{} & \times
    \left[ (N-2) A(\hat{s})^2 + \left( A(\hat{s}) + A(\hat{t})
      +A(\hat{u}) \right)^2 \right] \,.
\eqa
Here $\vp$ and $\vp^\prime$ denote the incoming and outgoing
momentum, respectively, $E(\vp)$ is the (lattice) energy of a
single Goldstone boson with momentum $\vp$ and
\be
    W = 2 E(\vp) = 2 E(\vp^\prime)
\eq
is the total energy. Furthermore,
\bea
     E^\prime (\vp) & = & \frac{\vp}{|\vp|} \cdot \nabla E(\vp)\,, \\
     A(z) & = & \frac{g_R}{3 m_R^2}
    \left[ \sinh^2 (m_R/2) + \om \sinh^2 m_R  \right]  z
  \nonumber \\ {} &{} & \times
    \left[ \sinh^2 (m_R/2) + \om \sinh^2 m_R  - \frac{1}{4}
       (1+4 \om)z \right] ^{-1} \,,   \\
  \hat{s} & = & \frac{4}{1+4\om} \left[ \sinh^2  E(\vp)
               + \om \sinh^2 \left(2 \flift E(\vp) \right) \right] \,, \\
  \hat{t} & = & -\frac{4}{1+4\om} \sum^3_{j=1}
    \left[ \sin^2 \left((p_j^\prime - p_j)/2 \right)
         + \om \sin^2 \left(p_j^\prime - p_j \right) \right]  \,, \\
  \hat{u} & = & -\frac{4}{1+4\om} \sum^3_{j=1}
    \left[ \sin^2 \left((p_j^\prime + p_j)/2 \right)
         + \om \sin^2 \left(p_j^\prime + p_j \right) \right]  \,.
\eqa
The lattice dispersion relation $E(\vp)$ is determined from the
(complex) poles of the free Goldstone boson propagator. For the
standard action ($\om =0$) it is given by
\be
     \cosh E(\vp) = 1 + \sum^3_{j=1} 2
                         \sin^2 \left( \frac{p_j}{2} \right)\,.
\eq
In the case of the improved action ($\om \!=\! -1/16$) or, more generally
for $\om \neq 0$, the situation is complicated by the appearance
of additional poles, which are unphysical in the sense that they
move to infinity in the continuum limit.  Moreover, for sufficiently
large momentum it can happen that there is no pole at all on the
imaginary $p_4$--axis.

If $ -1/4 < \om < 0$, one finds that a physical pole exists only
for momenta $\vp$ such that
\be
(1+4\om)^2 + 16\om \sum^3_{j=1} \left( \sin^2 (p_j/2)
+ \om \sin^2 p_j \right) \geq 0 \,.
\eq
%
\begin{figure}[th]
\vbox to 0mm{%
  \centerline{
      \setlength{\unitlength}{11mm}
      \begin{picture}(13,10)(0.00,0.00)
      \put(7.10,0.60){\makebox(0.0,0.0)[bc]{\large $m_R$}}
      \put(1.20,8.50){\makebox(0.0,0.0)[bc]{\Large $\delta$}}
      \end{picture}
      }\vss}
 \centerline{
 \epsfxsize=11cm
 \epsfbox{sfig9.ps}
 }
   \caption[The deviation $\dl$ between the lattice and continuum
      differential cross section for Goldstone boson scattering
      at $90^\circ$ versus the renormalized $\sg$--mass $m_R$.]{The
      deviation $\dl$ between the lattice and continuum
      differential cross section for Goldstone boson scattering
      at $90^\circ$ versus the renormalized $\sg$--mass $m_R$.
      The solid curves show $\dl$ in the case of the standard action
      and the dashed curves for the Symanzik improved action at fixed
      ratios $W/(2m_R) = 1.0$, $1.5$ and $2.0$ (from the right to the
      left).}
\lb{fig:cross}
\end{figure}

The corresponding dispersion relation is given by
\be
  \cosh E( \vp ) = - \frac{1}{4\om} + \frac{1}{4\om}
    \left[ (1+4\om)^2 + 16\om \sum^3_{j=1} \left( \sin^2 (p_j/2)
      + \om \sin^2 p_j \right)  \right]^{1/2}  \,.
\eq

We restrict ourselves to a scattering angle of $90^\circ$  with
$\vp$ and $\vp^\prime$ along coordinate axes. In fig.~\ref{fig:cross}
we plot the quantity
\be
            \dl = \frac{(d \sg / d \Omega)_{\mbox{\tiny latt.}}}
              {(d \sg / d \Omega)_{\mbox{\tiny cont.}}}  - 1
\eq
for $N\!=\!4$ as a function of $m_R$, where the ratio $R \!=\! W/(2 m_R)$ is
kept fixed at $R\!=\!1.0$, $1.5$, $2.0$.
One observes that the improved action improves on
$\dl$ only for sufficiently low values of $m_R$. For larger $m_R$,
it leads to even bigger cut-off effects than the standard action.
This behaviour might be due to the above mentioned unphysical poles.

We conclude that for reasonable choices of $R$ (e.g.\ $R=1.5$)
and of the limit on $\dl$ (5\% or 10\%, say) there is little
difference between the standard and the improved action as far as
the maximal admissible value of $m_R$ is concerned. For both actions one is
led to a number around 0.5. This then gives an upper bound on the
Higgs mass of $680 \, GeV$ for the standard action and of $750\, GeV$ for
the improved action.
These numbers should be compared with the results of Bhanot et~al.\
\cite{f4} obtained on an $F_4$--lattice. Applying in their case the same
criteria as in the case of the hypercubic lattice one arrives at an upper
bound of about $580 \, GeV$.

Another approach starts from the requirement that the upper bound,
if it is to be physically relevant at all, should be insensitive
to the specific regularization employed. So it should be read off
at a point where the curves corresponding to different regularizations
are sufficiently close to each other \cite{dejers}.
Depending on the precise
meaning of ``sufficiently close" it would follow from the observed
regularization dependence that the upper bound may be lower than
previously thought.
%
\section{Conclusions}
Let us summarize the main results of our investigation. We have
simulated the $O(4)$--symmetric $\Phi^4$--model in the broken phase
using the standard nearest-neighbour action as well as a tree--level
improved action \`{a} la Symanzik.
The configurations were generated by means of the reflection cluster
algorithm. The histogram method enabled us to continue our results to
$\kp$--values in the neighbourhood of those which were actually simulated.

Our main goal was a careful study of the scaling behaviour of the
model along with a determination of the triviality bound on the
Higgs mass and its regularization dependence. Due to the presence
of the massless Goldstone modes one has to cope with strong
finite--size effects and a reliable calculation of infinite--volume
quantities is possible only if these are kept under control. For
this purpose, chiral perturbation theory (at least to the order
worked out so far) is no longer adequate if the $\sg$--mass drops
below $\approx 0.5$, which is the case for most of our data points.
A limitation of this kind was, of course, to be expected. However,
if one is willing to apply renormalized perturbation theory
(thus assuming triviality of the model), it turns out that the
$1$--loop formulas provide a very satisfactory description
of the finite--size effects. Extracting the value of $m_\sg$
by means of perturbative formulas one gets the additional benefit
of bypassing the problems caused by the instability of the
$\sg$--particle: In perturbation theory the decay is automatically
taken into account.

  Here it is worthwhile to mention that at a $\sg$--correlation length
close to $2$ the results obtained by means of chiral perturbation theory
\cite{neuhaus} are completely consistent with those using renormalized
perturbation theory as described in the present paper.  Both ways of
analytically treating finite--size effects therefore share a common
region of applicability in the considered $O(4)$--model.

  As one approaches the critical point, the observables should obey mean
field scaling laws with logarithmic corrections.  We have studied the
scaling of the field expectation value $\Sg$ and the $\sg$--mass $m_\sg$.
In both cases we could verify the expectations.  There is, however, a
remarkable difference between the standard action and the improved
action.  In the case of the improved action the asymptotic form of the
scaling laws is already sufficient to describe the data.  For the data
from the standard action, on the other hand, the analogous fit leads to a
wrong exponent of the logarithmic correction, and a numerical integration
of the renormalization group equations is required in order to find
agreement with perturbation theory.

Combining our data on $\Sg$, $m_\sg$ and the Goldstone boson
wave function renormalization constant we could finally plot the
ratio $m_\sg/F$ versus $1/m_\sg$ for both actions. It is now almost a
matter of taste where to read off the upper bound on the $\sg$--mass,
and we leave it to the reader to choose his favourite point
(although for reasonable choices the answers will not differ very much).

\section*{Acknowledgement}
Helpful discussions with C.~Frick, K.~Jansen, J.~Jers\'{a}k,
and M.~L\"uscher are gratefully acknowledged.
Furthermore we wish to thank the HLRZ J\"ulich for providing the neccessary
computer time on its CRAY Y-MP.
%
\makeatletter
\if@double\else
\pagebreak[4]
\fi\makeatother
\section*{Appendices}
\appendix
\section{Saddle point approximation}
\lb{app:saddle}
We want to review the saddle point approximation for the integral
\be
       I = \int\limits_{0}^{\infty} \! dM
               F(M) \exp \left( \tlift - V {\cal U} (M) \right)\,,
\eq
as for example discussed in \cite{ding}
where $ {\cal U}(M) $ is supposed to be a real function in $ M $ which
has a unique minimum $ M_0 $.
We  consider the Taylor expansion for the function
$ {\cal U}(M) $ around $ M_0 $ and define a new integration
variable $ z(M) $ via
\bea
      \frac{1}{2} z^2 = {\cal U} (M) - {\cal U}(M_0)
                    & = & \frac{1}{2} (M-M_0)^2 {\cal U}'' (M_0)
                          + \sum_{n=3}^{\infty} \frac{1}{n!}
                              (M-M_0)^n \, {\cal U}^{(n)} (M_0) \,,\non \\
       {\cal U}'(M_0) = 0 & , & {\cal U}'' (M_0) > 0 \,.
\eqa
  With the substitution $ M \rightarrow z $,  the integral $I$ gets a
Gaussian form
\be
       I = \exp \left( \tlift - V{\cal U}(M_0) \right)
                \int\limits_{-\infty}^{+\infty} \! dz
                      \exp \left( - \frac{V}{2} z^2 \right)
                      \frac{F(M)}{ dz/dM } \,,
\eq
where $ z $ is taken as real along the path through the stationary point
$z=0$ and we are left to express $ F(M) $ and $ (dz/dM) $ in
terms of $ z $.

Since $ z \sim (M - M_0) $ for $ M \!\to\! M_0 $,  $ (M - M_0) $ can be
expanded  as a power series in $ z $ using Lagrange's reversion theorem
\cite{ding}
\be
      M  =  M_0 + \sum_{k=1}^{\infty} a_k z^k
         =  M_0 + \sum_{k=1}^{\infty} \left.
                  \left( \frac{d}{dM} \right)^{k-1}
                  \left( \frac{M-M_0}{z(M)} \right)^{k} \right|_{M=M_0}
                         \frac{z^k}{k!} \,.
\eq
The first few terms of the expansion of $ M $
are easy to calculate
\bea
      a_1 & = & \left( {\cal U}''(M_0) \right)^{-{1} / {2}} \,, \\
      a_2 & = & - \frac{1}{6} \frac{{\cal U}^{(3)}(M_0)}{
                  \left( \flift {\cal U}''(M_0) \right)^{2}} \,, \non\\
      a_3 & = & \frac{5}{72} \frac{
                  \left( {\cal U}^{(3)}(M_0) \right)^2 }{
                  \left( \flift {\cal U}''(M_0) \right)^{ {7} / {2}}}
                  - \frac{1}{24} \frac{ {\cal U}^{(4)}(M_0) }{
                     \left( \flift {\cal U}''(M_0) \right)^{ {5}/{2}}}
      \,. \non
\eqa

Now we only have to choose the function $ F(M) $ in order to
 extract the coefficients $ b_l $ of
 the power series for $F(M) (dM/dz) $ in $ z $
\be
      \frac{dM}{dz} F(M) =
                     \left( \sum_{k=1}^{\infty} k a_k z^{k-1} \right) \,\,
                     \left. F(M)\slift \right|_{
                     M= M_0 + \sum_{k=1}^{\infty} a_k z^k }
                         = \sum_{l=0}^{\infty} b_l z^l \,.
\eq

Finally using a basic integral formula we can calculate
the saddle point expansion of the integral $ I $
\bea
      I       & = & \exp \left( \tlift -V{\cal U}(M_0) \right)
                    \int\limits_{-\infty}^{+\infty} \! dz
                    \exp \left( - \frac{V}{2} z^2 \right)
                    \sum_{l=0}^{\infty} b_l z^l \non \\
      \mbox{} & = & \exp \left( \tlift -V{\cal U}(M_0) \right)
                    \sum_{l=0}^{\infty} b_{2l}
                    \frac{\Gm \left(\frac{1}{2} + l \right)}{
                          V^{ \frac{1}{2}+l }} 2^{ \frac{1}{2}+l }
                    z^l \,.
\lb{eq:saddle}
\eqa
Only $ b_l $ with even index contribute to this expansion.
Note that at lowest order
\be
      I        =  \exp \left( \tlift - V{\cal U}(M_0) \right)
                  \sqrt{ \frac{\ds 2 \pi}{\ds V {\cal U}'' (M_0)}} F (M_0)
                  \,.
\eq
%
\section{Integral representation
         of the finite volume correction $\bar{g}_r (m L)$}
\lb{app:g}
In this appendix we want to give an integral representation
for the finite--size correction (see eq.~(\ref{eq:gfunc}))
\bea
      \lb{eq:gfctapp}
      \bar{g}_r( m L) & = & L^{d-2r}
                          \left\{ \mlift \right. \frac{1}{L^d}
                           \sum_{ p\in \left( \frac{2\pi}{L} \ZZ \right)^d
                                }^{\hfill\prime\quad}
                            \left( p^2 + m^2 \right)^{-r}
                          - \int \!
                           \frac{d^d \! p}{(2\pi)^d} \,
                            \left( p^2 + m^2 \right)^{-r}
                          \left. \mlift \right\} \non \\
             \mbox{}  & = & \left\{ \mlift \right.
                           \sum_{ n\in \ZZ ^d }^{\hfill\prime\,}
                            \left( (2\pi n)^2 + y^2 \right)^{-r}
                          - \int \!
                          \frac{d^d \! z}{(2\pi)^d} \,
                            \left( z^2 + y^2 \right)^{-r}
                          \left. \mlift \right\}_{y=mL}
                           \,,
\eqa
which enters the perturbative expansions discussed in section
\ref{sec:finsiz} for $d=4$.
Note that $ \bar{g}_r $ is closely related to $g_r$
as defined by Hasenfratz and Leutwyler \cite{hale}:
\be
       g_r = \Gm (r) L^{2r-d} \left( \bar{g}_r (mL) + (mL)^{-2r} \right)\,.
\eq

We review some tools we need for further calculations.
\begin{itemize}
\item
   The Poisson formula connects sums on dual lattices
   \be
         L^{-d} \sum_{ p\in \left( \frac{2\pi}{L} \ZZ \right)^d } F(p)
              = \sum_{ x \in \left( L \ZZ \right)^d } \tl{F}(x)
   \lb{eq:poiss}
   \eq
   where $ \tl{F}(x) $ is the Fourier transform of $ F(p) $
   \be
         \tl{F} (x) = \int \!
         \frac{d^d \! p}{(2\pi)^d} \;
                           e^{ipx} \; F(p) \,.
   \lb{eq:four}
   \eq
\item
   The sum
   \be
         \theta (\tau) = \sum_{n \in \ZZ^d} e^{- \pi \tau n^2}
   \eq
   is a special case of the so--called $ \theta $--function.
   Under the special modular transformation $ \tau \!\to\! 1/\tau $
   it behaves according to the relation
   \be
         \theta (\tau) = \left(\frac{1}{\sqrt{\tau}}\right)^d \;
                      \theta \left(\frac{1}{\tau}\right) \,,
   \eq
   which can be proved using the Poisson formula.
\item
   We will later use the Fourier transformation
   \be
              \int \!
              \frac{d^d \! p}{(2\pi)^d} \;
              e^{ipx} \; \exp\left(  -\frac{1}{2} p^T B p \right)
            = \left[ 2\pi \det B \right]^{-d/2} \,
              \exp\left( -\frac{1}{2} x^T B^{-1} x \right)
   \lb{eq:fouri}
   \eq
   where $ B $ is supposed to be a symmetric, positive definite matrix.
   Furthermore we need the simple Laplace transformation
   \be
         \frac{1}{\left( p^2 + m^2 \right)^{r} }
         =   \int\limits_{0}^{\infty} \! d\lm \,
             \frac{\lm^{r-1}}{\Gm (r) } \,
             \exp \left\{ \slift - \lm \left(p^2 + m^2 \right) \right\}
             \quad \quad \left(\Real (p^2 + m^2 ) > 0\right) \,.
   \lb{eq:lapl}
\eq
\end{itemize}

We now apply the Poisson summation formula
for $ F(p) \!=\! \left( p^2+m^2\right )^{-r}$:
By means of the (\ref{eq:lapl}) we get
\be
      \tl{F}(x) = \int\limits_{0}^{\infty} \! d\lm \,
                  \frac{\lm^{r-1}}{\Gm (r) } \,
                  e^{ - \lm m^2 }
                  \int \!
                  \frac{d^d \! p}{(2\pi)^d} \;
                  e^{ipx} \; e^{ - \lm p^2 } \,,
\eq
and
evaluating the Fourier transformation with the help eq.~(\ref{eq:fouri})
we obtain
\be
      \frac{1}{L^d} \sum_{ p\in \left( \frac{2\pi}{L} \ZZ \right)^d}
      \left( p^2 + m^2 \right)^{-r}
      =   \sum_{ x \in \left( L \ZZ \right)^d}
          \int\limits_{0}^{\infty} \! d\lm \,
          \frac{\lm^{r-1}}{\Gm (r) } \,
          \left( 4 \pi \lm \right)^{-d/2}
          \exp \left( - \lm m^2 - \frac{x^2}{4 \lm} \right) \,.
\eq
In order to avoid a possible singularity at $ \lm \!=\! 0 $
we have to separate the contribution from $ x \!=\! 0 $:
\bea
\lb{eq:latprop}
      \frac{1}{L^d} \sum_{ p\in \left( \frac{2\pi}{L} \ZZ \right)^d}
      \left( p^2 + m^2 \right)^{-r}
              & = & \int\!
                    \frac{d^d \! p}{(2\pi)^d}
                    \left( p^2 + m^2 \right)^{-r}  \\
      \mbox{} &   & + \int\limits_{0}^{\infty} \! d\lm \,
                      \frac{\lm^{r-1}}{\Gm (r) } \,
                      \left( 4 \pi \lm \right)^{-d/2}
                       e^{- \lm m^2 }
                       \sum_{x \in \left( L \ZZ \right)^d
                             }^{\hfill\prime \;\;\;}
                        \exp \left( - \frac{x^2}{4 \lm} \right)
\non \,.
\eqa
Multiplying eq.~(\ref{eq:latprop}) by $L^{d-2r}$
we get the following integral representation for the finite--size
correction:
\be
      \bar{g}_r(m L) = \int\limits_{0}^{\infty} \! d \tau \,
                               \frac{ \tau^{r-1}}{\Gm (r) } \,
                               \left( 4 \pi \tau \right)^{-d/2}
                               e^{- \tau \left( m L \right)^2 }
                               \sum_{n \in \ZZ^d
                                       }^{\hfill\prime\,}
                               \exp \left( - \frac{n^2}{4 \tau} \right) \;
                             - \frac{1}{\left( m L \right)^{2r} } \,.
\eq

For the numerical computation
we divide the integral into three parts:
\be
      \bar{g}_r(m L)  = \left( 4 \pi \right)^{-r} \Gm (r) ^{-1}
                           \left( a_r (x) + b_r (x) - b_{r-d/2} (x)
                           - x^{-r} \Gm (r)
                           \right)_{x = \frac{\left( m L \right)^2 }{4\pi}}
                           \,.
\eq
The function $ b_s (x) $ may be expressed in terms  of the incomplete
 $ \Gm $ function,
\be
      b_s (x) = x^{-s} \Gm \left( s , x \right)
              = x^{-s} \int\limits_{x}^{\infty} \! d \lm \,
                       \lm^{s-1} \, e^{- \lm x} \,,
\eq
and $ a_s (x) $ has a convergent integral representation
\be
      a_s (x) = \int\limits_{0}^{1} \! d\tau \,
                \left( \tau^{s-1-d/2} \, e^{- \tau x }
                       + \tau^{-s-1} \, e^{- x / \tau }
                \right)
                \left[ \theta \left( \frac{1}{\tau} \right) - 1 \right]\,,
\eq
which can be evaluated numerically by  Chebyshev integration.

\end{document}